\documentclass[12pt,aps,prb,showpacs,amsmath,amssymb,superscriptaddress,floatfix]{revtex4-2}

\usepackage[english]{babel}
\usepackage[utf8]{inputenc}
\usepackage{graphicx}
\usepackage{bm} 
\usepackage{latexsym}
\usepackage{amsfonts}
\usepackage{bbold}
\usepackage{appendix}
\usepackage[colorlinks = true,
            linkcolor = blue,
            urlcolor  = blue,
            citecolor = blue,
            anchorcolor = blue]{hyperref}
\usepackage{mathdots}
\newcommand{\eps}{\epsilon}
\newcommand{\sg}{\sigma}
\newcommand{\Sg}{\Sigma}
\newcommand{\alf}{\alpha}
\newcommand{\gam}{\gamma}
\newcommand{\Gam}{\Gamma}
\newcommand{\lam}{\lambda}
\newcommand{\Lam}{\Lambda}

\newcommand{\up}{\uparrow}
\newcommand{\down}{\downarrow}

\newcommand{\bA}{\mathbf{A}}
\newcommand{\bh}{\mathbf{h}}
\newcommand{\bk}{\mathbf{k}}
\newcommand{\bn}{\mathbf{n}}
\newcommand{\bq}{\mathbf{q}}
\newcommand{\bQ}{\mathbf{Q}}
\newcommand{\br}{\mathbf{r}}

\newcommand{\bS}{\mathbf{S}}

\newcommand{\cB}{\mathcal{B}}
\newcommand{\cG}{\mathcal{G}}
\newcommand{\cH}{\mathcal{H}}
\newcommand{\cJ}{\mathcal{J}}

\newcommand{\cP}{\mathcal{P}}
\newcommand{\cR}{\mathcal{R}}
\newcommand{\cS}{\mathcal{S}}
\newcommand{\cT}{\mathcal{T}}
\newcommand{\cU}{\mathcal{U}}

\newcommand{\Ct}{{\widetilde C}}
\newcommand{\chit}{{\widetilde\chi}}

\newcommand{\tr}{{\rm tr}}

\newcommand{\bra}{\langle}
\newcommand{\ket}{\rangle}

\begin{document}

\author{Henrik M\"uller-Groeling}
\email{h.mueller-groeling@fkf.mpg.de}
\affiliation{Max Planck Institute for Solid State Research, Heisenbergstrasse 1, D-70569 Stuttgart, Germany}

\author{Pietro M.\ Bonetti}
\affiliation{Department of Physics, Harvard University, Cambridge MA 02138, USA}
\affiliation{Max Planck Institute for Solid State Research, Heisenbergstrasse 1, D-70569 Stuttgart, Germany}

\author{Paulo Forni}
\affiliation{Max Planck Institute for Solid State Research, Heisenbergstrasse 1, D-70569 Stuttgart, Germany}

\author{Walter Metzner}
\email{w.metzner@fkf.mpg.de}
\affiliation{Max Planck Institute for Solid State Research, Heisenbergstrasse 1, D-70569 Stuttgart, Germany}

\title{SU(2) gauge theory of fluctuating stripe order in the two-dimensional Hubbard model}

\date{\today}

\begin{abstract}
We present an SU(2) gauge theory of fluctuating stripe order in the two-dimensional Hubbard model. The theory is based on a fractionalization of the electron operators in fermionic chargons with a pseudospin degree of freedom, and charge neutral spinons capturing fluctuations of the spin orientation. The chargons are treated in a renormalized mean-field theory. We focus on regions of the phase diagram where they undergo stripe order. The spinons are described by a non-linear sigma model with pseudospin stiffnesses determined by the chargons. They prevent breaking of the physical SU(2) spin symmetry at any finite temperature, resulting in a charge ordered pseudogap phase with a reconstructed Fermi surface and a spin gap. The spectral function for single-particle excitations exhibits a collection of Fermi arcs and other structures. The arcs appear in various regions of the Brillouin zone, but never exclusively near the Brillouin zone diagonals.
For a suitable choice of the bare dispersion relation, the form of the low energy spectral function resembles experimental observations from angular resolved photoemission on cuprates in which stripe fluctuations play an important role.
\end{abstract}

\maketitle


\section{Introduction}

A universal hallmark of cuprate superconductors -- in addition to their high superconducting transition temperatures -- is their pseudogap behavior above $T_c$, which has been observed in a broad hole-doping range from the underdoped into the optimally doped regime \cite{Keimer2015}.
The pseudogap phenomenology includes several striking properties: a reduction of the charge carrier concentration, a spin gap, and a reconstructed Fermi surface which in photoemission looks like a collection of Fermi arcs. It is often accompanied by electronic nematicy, breaking the tetragonal symmetry of the crystal, and a tendency toward charge order. Suppressing superconductivity by a high magnetic field, the pseudogap regime extends into the otherwise superconducting region of the phase diagram -- for doping concentrations as high as 20 percent \cite{Proust2019}.

A minimal model describing the strongly correlated valence electrons in the copperoxide planes of cuprate superconductors is the two-dimensional Hubbard model \cite{Anderson1987, Zhang1988}.
There is numerical evidence for pseudogap behavior in the two-dimensional Hubbard model, in particular from quantum cluster calculations \cite{Qin2022}, and a careful analysis of the contributions to the self-energy from effective charge, spin, and pairing interactions has revealed that the pseudogap is generated mainly by antiferromagnetic spin fluctuations \cite{Gunnarsson2015}.
Approximate theories based on quantum field theoretic methods can provide further insight into the structure and mechanisms of the pseudogap behavior, and results for spectroscopic observables with a high energy and momentum resolution. Early theories of the pseudogap phenomenon were based on weak-coupling expansions, such as the Moriya theory \cite{Moriya2000} and the two-particle self-consistent approach \cite{Vilk1996}. However, the pseudogap forms only for large magnetic correlation lengths and thus in a narrow parameter regime in these theories, while numerical results show that strong short-ranged correlations are actually sufficient.
Alternative previous theories involving only short-range order are based on Anderson's resonanting valence bond (RVB) concept \cite{Lee2006}, and on dynamical gap generation by umklapp processes \cite{Yang2006}.

More recently it was shown that the pseudogap behavior observed in cuprates can be captured by SU(2) gauge theories of fluctuating magnetic order \cite{Sachdev2016, Chatterjee2017, Scheurer2018, Wu2018, Sachdev2019, Bonetti2022a, Forni2025}. This approach is based on a fractionalization of the electron into a fermionic {\em chargon}\/ with a pseudospin degree of freedom and a charge neutral {\em spinon}, that is, an SU(2) matrix providing a space and time dependent fluctuating local reference frame \cite{Schulz1995}. The local spin rotations can be parametrized by an SU(2) gauge field. One can then construct solutions where the chargons exhibit some sort of magnetic order (in particular, N\'eel, spiral, or stripe), and the Fermi surface gets correspondingly reconstructed. The spinon fluctuations prevent magnetic long-range order of the physical spin-carrying electrons at finite temperatures, and, if strong enough, also in the ground state \cite{Dupuis2002, Borejsza2004, Sachdev2009}. Quantities involving only charge degrees of freedom behave as in a conventional magnetically ordered state \cite{Eberlein2016, Mitscherling2018, Bonetti2020}.
Recently a more sophisticated gauge theory of the pseudogap regime with additional auxiliary degrees of freedom (``ancilla qubits'') was proposed \cite{Zhang2020, Zhang2020a, Mascot2022, Nikolaenko2023}.

So far, concrete evaluations of SU(2) gauge theories of fluctuating magnetism were restricted to N\'eel or planar spiral order. In both cases the magnetic order leads to a splitting of the bare band structure into (only) two quasi-particle bands. In the presence of a sizable next-to-nearest neighbor hopping, as present in the cuprates, N\'eel order applies to the electron-doped, and spiral order to the lightly hole-doped Hubbard model. However, at larger hole-doping stripe order becomes favorable, at least in mean-field theory \cite{Scholle2023, Scholle2024}, and several numerical simulations of the Hubbard model with a sizable hole-doping indicate a stripe-ordered ground state \cite{Zheng2017, Huang2018, Qin2020, Xu2022}, sometimes coexisting with superconductivity \cite{Xu2024}.

In this paper we formulate and evaluate an SU(2) gauge theory of fluctuating stripe order. Our analysis extends the formalism of Ref.~\cite{Bonetti2022a}, which allows for a (approximate) microscopic calculation of observables for interacting electron models.
Fluctuating stripe states have also been considered recently in Refs.~\cite{Schloemer2025, Zhang2025}.
The main complication for stripe order, compared to N\'eel or spiral order, is the large number of quasi-particle bands, which is proportional to the periodicity of the stripe pattern. We solve the chargon sector in a renormalized mean-field theory, and we compute the pseudospin stiffnesses from a renormalized random phase approximation (RPA). The spinon dynamics is governed by a nonlinear sigma model, which is solved in a saddle point approximation. We present results for the quasi-particle Fermi surfaces, the stiffnesses, and the electron spectral function.

The remainder of the article is structured as follows. In Sec.~II we describe the general structure of the SU(2) gauge theory of fluctuating stripe order, and our approximate solutions for the chargon and spinon dynamics. Concrete results for the two-dimensional Hubbard model are presented in Sec.~III. A conclusion in Sec.~IV closes the presentation.


\section{Model and Method}
\label{sec: method}

While our method could be used for a broader class of models, we consider the Hubbard model on a square lattice \cite{Arovas2022, Qin2022} for the sake of concreteness. Our calculations are based on the functional integral formalism for quantum many-particle systems \cite{Negele1987}. The Hubbard action in imaginary time reads as
\begin{eqnarray} \label{eq: Hubbard action}
 \mathcal{S}[c,c^*] &=&
 \int_0^\beta\!d\tau \sum_{j,j',\sigma} c_{j\sg}^*
 \left[ \left( \partial_\tau - \mu\right) \delta_{jj'} + t_{jj'} \right] c_{j'\sg}
 \nonumber \\
 &+& \int_0^\beta\!d\tau \; U \sum_j n_{j\up}n_{j\down} ,
\end{eqnarray}
where $c_{j\sg} = c_{j\sg}(\tau)$ and $c_{j\sg}^* = c_{j\sg}^*(\tau)$ are Grassmann fields corresponding to the annihilation and creation, respectively, of an electron with spin orientation $\sg$ at site $j$, and
$n_{j\sg} = n_{j\sg}(\tau) = c_{j\sg}^*(\tau) c_{j\sg}(\tau)$.
The hopping amplitudes $t_{jj'}$ are translation invariant, that is, they depend only on $j-j'$.
The chemical potential is denoted by $\mu$, and $U > 0$ is the strength of the (repulsive) Hubbard interaction. To simplify the notation, we write the dependence of the fields on the imaginary time $\tau$ only if needed for clarity.
The Hubbard action in \eqref{eq: Hubbard action} is invariant under \emph{global} SU(2) spin rotations $c_j(\tau) \to \mathcal{U} c_j(\tau)$,
$c_j^*(\tau) \to c_j^*(\tau) \, \mathcal{U}^\dag$,
where $c_j$ and $c^*_j$ are two-component spinors composed from $c_{j\sg}$ and $c^*_{j\sg}$ with $\sg \in \{\up,\down\}$, respectively, while $\mathcal{U}$ is an arbitrary SU(2) matrix.


\subsection{Electron fractionalization}

To separate spin orientation fluctuations from the charge degrees of freedom, we fractionalize the electronic fields as~\cite{Schulz1995, Dupuis2002, Borejsza2004, Sachdev2009}
\begin{equation} \label{eq: electron fract}
 c_j(\tau) = R_j(\tau) \, \psi_j(\tau)  \, , \quad
 c_j^*(\tau) = \psi_j^*(\tau) \, R_j^\dag(\tau) ,
\end{equation}
where $R_j(\tau) \in \mbox{SU(2)}$, to which we refer as {\em spinon}\/, is composed of bosonic fields, and the components $\psi_{js}$ of the {\em chargon}\/ spinor $\psi_j$ are fermionic. The spinons transform under the global SU(2) spin rotation by a \emph{left} matrix multiplication, while the chargons are left invariant. Conversely, a U(1) charge transformation acts only on $\psi_j$, leaving $R_j$ unaffected.
The factorization in Eq.~\eqref{eq: electron fract} introduces a redundant {\em local}\/ SU(2) symmetry, acting as
\begin{subequations} \label{eq: gauge symmetry}
\begin{align}
 & \psi_j(\tau) \to \mathcal{V}_j(\tau) \, \psi_j(\tau) , \quad
   \psi^*_j(\tau) \to \psi^*_j(\tau) \, \mathcal{V}^\dag_j(\tau) , \\
 & R_j(\tau) \to R_j(\tau) \, \mathcal{V}_j^\dag(\tau), \quad\!\!
   R_j^\dag(\tau) \to \mathcal{V}_j(\tau) \, R_j^\dag(\tau) \, ,
\end{align}
\end{subequations}
with $\mathcal{V}_j\in \mbox{SU(2)}$.
Hence, the components $\psi_{js}$ of $\psi_j$ carry an SU(2) gauge index $s$, while the components $R_{j,\sg s}$ of $R_j$ have two indices, the first one ($\sg$) corresponding to the global SU(2) symmetry, and the second one ($s$) to SU(2) gauge transformations.

We now rewrite the Hubbard action in terms of the spinon and chargon fields. The quadratic part of \eqref{eq: Hubbard action} can be expressed as \cite{Borejsza2004}
\begin{eqnarray} \label{eq: S0 chargons spinons}
 \mathcal{S}_0[\psi,\psi^*,R] &=& \int_0^\beta\!d\tau
 \bigg\{ \sum_j \psi^*_j \left[ \partial_\tau - \mu - \phi_j \right]
 \psi_{j} \nonumber \\
 && + \, \sum_{j,j'} t_{jj'} \, \psi_j^* \, e^{iA_{jj'}} \, \psi_{j'} \bigg\},
\end{eqnarray}
where we have introduced a ficticious SU(2) gauge field, defined by
\begin{equation} \label{eq: gauge field def}
\begin{split}
 \phi_j(\tau) &= - R_j^\dag(\tau) \partial_\tau R_j(\tau) \, , \\
  e^{iA_{jj'}(\tau)} &= R_j^\dag(\tau) R_{j'}(\tau) \, .
\end{split}
\end{equation}
Both $\phi_j(\tau)$ and $A_{jj'}(\tau)$ are elements of the Lie algebra of SU(2), and can therefore be written as linear combinations of the Pauli matrices $\sg^1$, $\sg^2$, and $\sg^3$:
\begin{equation} \label{eq: lie algebra}
\begin{split}
 \phi_j(\tau) &= \frac{1}{2} \sum_a \phi_j^a(\tau) \sigma^a \, , \\
  A_{jj'}(\tau) &= \frac{1}{2} \sum_a A_{jj'}^a(\tau) \sigma^a \, .
\end{split}
\end{equation}
$A_{jj'}(\tau)$ contributes to the action only if the sites $j$ and $j'$ are coupled by a hopping amplitude.

In the interaction part of the Hubbard action \eqref{eq: Hubbard action}, the SU(2) rotations drop out \cite{Bonetti2022a}, so that
\begin{equation} \label{eq: S_int}
 \mathcal{S}_\mathrm{int}[\psi,\psi^*,R] = \mathcal{S}_\mathrm{int}[\psi,\psi^*] =
 \int_0^\beta\!d\tau \, U \sum_j n^\psi_{j\up}n^\psi_{j\down},
\end{equation}
with $n^\psi_{js} = \psi^*_{js} \psi_{js}$.
The action $\mathcal{S} = \mathcal{S}_0 + \mathcal{S}_\mathrm{int}$ is nothing but the Hubbard model action where the physical electrons have been replaced by chargons coupled to an SU(2) gauge field.


\subsection{Spinon effective action}

We now derive an effective action for the spinon fields. Assigning only long wave length spin orientation fluctuations to the rotation matrices $R_j(\tau)$, we can assume that $R_j(\tau)$ is slowly varying in space. Under this assumption, the gauge fields $A_{jj'}^a(\tau)$ can be linearized as
\begin{equation}
 A_{jj'}^a(\tau) = \br_{jj'} \cdot \bA_j^a(\tau) =
 x_{jj'} A_{x,j}^a(\tau) + y_{jj'} A_{y,j}^a(\tau) \, ,
\end{equation}
where $\br_{jj'} = \br_j - \br_{j'}$ is the real space vector from site $j'$ to site $j$, while $x_{jj'}$ and $y_{jj'}$ are its $x$ and $y$ components, respectively. The spatial components of the gauge field $A_{\alf,j}^a$ with $\alf=x,y$ can be combined with the temporal components $A_{0,j}^a \equiv \phi_j^a$ to a 3-vector $A_{\mu,j}^a$ with $\mu=0,x,y$.

For slowly varying rotation matrices $R_j(\tau)$, an expansion in the gauge fields corresponds to a gradient expansion. We will carry out this expansion to quadratic order. Hence, we derive the effective action for the spinons from an expansion of $\cS_0[\psi,\psi^*,A]$ in Eq.~\eqref{eq: S0 chargons spinons} to quadratic order in $A$.
Fourier transforming $\psi_j(\tau)$ to $\psi(k)$ with $k = (\bk,ik_0) = (k_x,k_y,ik_0)$, and $A_{\mu,j}^a(\tau)$ to $A_\mu^a(q)$ with $q = (\bq,iq_0) = (q_x,q_y,iq_0)$,
one obtains \cite{Bonetti2022ward}
\begin{equation} \label{eq: S_0 expanded}
\begin{split}
 \cS_0[\psi,\psi^*,A] &=
 - \int_k \psi^*(k) \left( ik_0 + \mu - \eps_\bk \right) \psi(k) \\
 &- \frac{1}{2} \int_{k,q} \sum_a A_\mu^a(q) \gam_\bk^\mu \, \psi^*(k+q) \sg^a \psi(k) \\
 &+ \frac{1}{8} \int_{k,q,q'} \sum_a A_\alf^a(q-q') A_\beta^a(q') \gam_\bk^{\alf\beta} \,
 \psi^*(k+q) \psi(k) \, ,
\end{split}
\end{equation}
with the first and second order coupling functions $\gam_\bk^\mu = (1,\nabla_\bk \eps_\bk)$ and $\gam_\bk^{\alf\beta} = \partial_{k_\alf k_\beta}^2 \eps_\bk$, respectively, and $\eps_\bk$ is the Fourier transform of the hopping amplitudes $t_{jj'}$. Here and in the following equations, we use Einstein's summation convention for repeated space-time indices, while sums over spin indices are written explicitly.
The integrals are written in a shorthand notation
$\int_k = T \sum_{k_0} \int\frac{d^2\bk}{(2\pi)^2}$, and analogously for $\int_q$, where $k_0$ and $q_0$ are fermionic and bosonic Matsubara frequencies, respectively.

Integrating out the fermionic fields from the action
$\cS[\psi,\psi^*,A] = \cS_0[\psi,\psi^*,A] + \mathcal{S}_\mathrm{int}[\psi,\psi^*]$, one obtains an effective action $\cS[A]$ for the gauge fields. For slowly varying fluctuations, we can expand this action in powers of $A$. The linear term vanishes \cite{Bonetti2022a} except for a Berry term \cite{Borejsza2004} which is probably irrelevant \cite{Auerbach1994}. The quadratic term has the form
\begin{equation} \label{eq: eff action general}
 \cS[A] = \frac{1}{2} \int_{q,q'} \sum_{a,b} K_{\mu\nu}^{ab}(q,q') \,
 A_\mu^a(q) \, A_\nu^b(-q') \, ,
\end{equation}
with a function $K_{\mu\nu}^{ab}$ we refer to as ``gauge kernel'' in the following.
While the frequencies $q_0$ and $q'_0$ contributing to Eq.~\eqref{eq: eff action general} are always equal, the momenta $\bq$ and $\bq'$ can differ by reciprocal lattice vectors of the magnetic state, which breaks the translation invariance. If $\bq$ and $\bq'$ are so small that $\bq - \bq'$ is smaller than any non-zero reciprocal lattice vector, one can assume that $\bq = \bq'$, and Eq.~\eqref{eq: eff action general} simplifies to
\begin{equation} \label{eq: eff action}
 \cS[A] = \frac{1}{2} \int_q \sum_{a,b} K_{\mu\nu}^{ab}(q) \,
 A_\mu^a(q) \, A_\nu^b(-q) \, .
\end{equation}
In the following we will consider only the contributions from $\bq = \bq'$.
The gauge kernel is a sum of a {\em paramagnetic}\/ and a {\em diamagnetic} contribution, $K_{\mu\nu}^{ab} = K_{\mu\nu}^{p,ab} + K_{\mu\nu}^{d,ab}$, where
\begin{subequations} \label{eq: gauge kernel}
\begin{align}
 K_{\mu\nu}^{p,ab}(q) &=
 - \frac{1}{4} \int_{k,k'} \gam_\bk^\mu \gam_{\bk'}^\nu \,
 \bra \psi^*(k-q) \sg^a \psi(k) \, \psi^*(k'+q) \sg^b \psi(k') \ket \, , \\
 K_{\mu\nu}^{d,ab} &=
 \frac{1}{4} \delta_{ab} \delta_{\mu\alf} \delta_{\nu\beta} \int_k
 \gam_\bk^{\alf\beta} \bra \psi^*(k) \psi(k) \ket \, .
\end{align}
\end{subequations}
The brackets $\bra...\ket$ are connected expectation values for chargons in the absence of the gauge field $A$, but in the presence of an infinitesimal SU(2) symmetry breaking external field $h$, whose role and precise form will be described later.
The diamagnetic term is momentum and frequency independent, and contributes only if $\mu$ and $\nu$ are both spatial indices $\alf,\beta \in \{x,y\}$.

In this work we approximate the gauge kernel $K_{\mu\nu}^{ab}(q)$ in the effective action $\cS[A]$ by its low energy and long wavelength limit $q \to 0$.
More specifically, its spatial components $K_{\alpha\beta}^{ab}(q)$ with $\alpha,\beta \in \{x,y\}$ are approximated by the {\em spatial spin stiffness}
\begin{equation} \label{eq: spatial stiffness}
 J_{\alf\beta}^{ab} =
 \lim_{h \to 0} \lim_{\bq \to {\bf 0}} K_{\alf\beta}^{ab}(\bq,0) \, ,
\end{equation}
and its temporal component $K_{00}^{ab}(q)$ is approximated by the {\em temporal spin stiffness}
\begin{equation} \label{eq: temporal stiffness}
 Z^{ab} = - J_{00}^{ab} =
 - \lim_{h \to 0} \lim_{q_0 \to 0} K_{00}^{ab}({\bf 0},iq_0) \, .
\end{equation}
The discrete set of Matsubara frequencies at finite temperature has been continued to the entire imaginary frequency axis to take the limit $q_0 \to 0$. Alternatively one may analytically continue $q_0$ to the real frequency axis, and then take the zero frequency limit along that axis. The limits $\bq \to {\bf 0}$, $q_0 \to 0$, and $h \to 0$ do not commute. Their order must therefore be specified.
We neglect the mixed spatio-temporal components $J_{\alpha0}^{ab}$ and $J_{0\beta}^{ab}$. They are imaginary and due to Landau damping.

The action in Eq.~\eqref{eq: eff action} can then be written as a {\em local}\/ action in a real space continuum representation,
\begin{equation} \label{eq: local action}
 \cS[A] = \frac{1}{2} \int dx \sum_{a,b} J_{\mu\nu}^{ab} \, A_\mu^a(x) A_\nu^b(x) \, ,
\end{equation}
where $x = (\br,-i\tau)$ and $A_\mu^a(x)$ is the Fourier transform of $A_\mu^a(q)$.
This form of the effective action has already been used in the case of spiral magnetic order in \cite{Bonetti2022a}.
The action \eqref{eq: local action} can be mapped to a nonlinear sigma model of the form \cite{Bonetti2022a}
\begin{equation} \label{eq: nlsm}
 \cS[\cR] = \frac{1}{2} \int dx \, \tr \left\{ \cP_{\mu\nu}
 [\partial_\mu \cR^T(x)] [\partial_\nu \cR(x)] \right\} \, ,
\end{equation}
where $\cR$ is the adjoint representation of the SU(2) rotation $R$, defined by
\begin{equation}
 R^\dag \sg^a R = \sum_b \cR^{ab} \sg^b \, ,
\end{equation}
and $\partial_\mu = (\nabla,i\partial_\tau)$. The coefficients are given by
$\cP_{\mu\nu} = \frac{1}{2} \tr(\cJ_{\mu\nu}) \mathbb{1} - \cJ_{\mu\nu}$, where
$\cJ_{\mu\nu}$ is the $3\times3$ stiffness matrix with matrix elements $J_{\mu\nu}^{ab}$.


\subsection{Chargons} \label{sec: chargons}


\subsubsection{Renormalized mean-field theory} \label{sec: fRG}

We deal with the chargons in a renormalized mean-field theory as formulated in Ref.~\cite{Wang2014}. Mean-field equations for ordered phases are thereby solved with renormalized instead of bare interactions as input parameters. The renormalized interactions are obtained from a functional renormalization group flow \cite{Metzner2012}, which takes charge, spin, and pairing channels into account on equal footing.
This approach has been applied to the Hubbard model in several papers \cite{Yamase2016, Vilardi2020, Bonetti2022a}. The coupled fluctuations of the various channels have two important effects: i) magnetic interactions are reduced, and ii) d-wave pairing interactions are generated by fluctuations \cite{Metzner2012}.
Since we are interested in the normal (non-superconducting) state, we ignore the pairing instability by assuming that the temperature is above the transition temperature for pairing, or that superconductivity has been suppressed by an external magnetic field.

The effective interactions are momentum and frequency dependent, but these dependences are rather weak for the {\em two-particle irreducible}\/ \cite{Wang2014} effective interactions which enter the mean-field equations. Hence we compute the effective interaction in the magnetic channel at zero frequency and at the wave vector $\bQ$ where the magnetic interaction is maximal, and use this as a momentum and frequency independent coupling constant $\bar U$.
Using the most general mean-field (Hartree-Fock) decoupling of the effective interaction in the charge and spin channels \cite{Zaanen1989, Scholle2023}, one then obtains the renormalized mean-field interaction
\begin{equation}\label{eq: MF decoupling}
\begin{split}
 H_{\rm MF}^I &= \sum_{j} \left
 [\Delta^c_j \, \psi^\dagger_j \sg^0 \psi_j - {\bar U}^{-1} (\Delta_j^c)^2 \right] \\
 &- \sum_{j} \sum_{a=1}^3 \left
 [\Delta^a_j \, \psi^\dagger_j \sg^a \psi_j - {\bar U}^{-1} (\Delta_j^a)^2 \right] \, ,
\end{split}
\end{equation}
where $\psi_j$ is the two-component spinor composed of the annihilation operators $\psi_{j\up}$ and $\psi_{j\down}$, and $\psi_j^\dag$ is its hermitian conjugate; $\sigma^0$ is the two-dimensional identity matrix and $\sigma^1,\sigma^2,\sigma^3$ are the Pauli matrices.


\subsubsection{Stripe order} \label{sec: stripe order}

In a sizable parameter range, the energetically most favorable solution of the (renormalized) mean-field equations yields collinear spin order accompanied by charge order, also known as spin-charge stripe order \cite{Schulz1989, Zaanen1989, Machida1989, Poilblanc1989, Schulz1990, Kato1990, Scholle2023}. In the following we focus our analysis on this stripe-ordered regime.
The direction of the collinear spin order can be chosen arbitrarily. Assuming an orientation of the spins in $x$-direction, the order parameters $\Delta_j^2$ and $\Delta_j^3$ in Eq.~\eqref{eq: MF decoupling} vanish, and we are left with $\Delta_j^c$ and $\Delta_j^s \equiv \Delta_j^1$ defined by the expectation values
\begin{subequations} \label{eq: gaps}
\begin{align}
 \Delta^c_j &= \frac{1}{2} {\bar U} \langle \psi_j^\dag \sg^0 \psi_j \rangle =
 \frac{1}{2} {\bar U} \, \rho_j \, , \\
 \Delta^s_j &= \frac{1}{2} {\bar U} \langle \psi_j^\dag \sg^1 \psi_j \rangle =
 {\bar U} \, m_j^1 \, ,
\end{align}
\end{subequations}
where $\rho_j$ is the local charge density, and $m_j^1$ the local spin density in $x$-direction.
Except for a restricted region near the edge of the magnetically ordered regime (at fairly large doping), the stripe order is usually {\em unidirectional}, that is, along one of the axes of the square lattice spins order perfectly antiferromagnetically, while the charge density is uniform along that axis \cite{Scholle2023, Scholle2024}. Assuming an antiferromagnetic alignment (and uniform charge density) in the $y$-direction, the wave vectors describing the stripe order in Fourier space have the form $\bQ = (\pi - 2\pi\eta,\pi)$, where $\eta$ parametrizes the deviation from N\'eel order in $x$-direction.

For concrete calculations we assume that the stripe order is commensurate with a period $p$, that is, the ordering pattern is invariant under translations by $p$ lattice sites in $x$-direction. Incommensurate wave vectors can be approximated to any desired accuracy by commensurate wave vectors with sufficiently large periods $p$.
The spin and charge profile can then be expanded in a finite Fourier series \cite{Scholle2024}
\begin{subequations} \label{eq: profiles}
\begin{align}
 m_j^1 &= \sum_{n \, \text{odd}} \, M_n \, e^{i \bQ_n \cdot \mathbf{r}_j} \, , \\
 \rho_j &= \sum_{n \, \text{even}} \varrho_n \, e^{i \bQ_n \cdot \mathbf{r}_j} \, ,
\end{align}
\end{subequations}
where $\bQ_n = (2\pi n/p,\pi n)$, and $\mathbf{r}_j$ are the coordinates of the lattice site $j$. The indices $n$ are integer numbers from $\{ 0,1,\dots,P-1 \}$, where $P$ is the smallest positive integer satisfying $P \bQ_1 = (0,0)$ modulo reciprocal lattice vectors. Hence, $P=p$ if $p$ is even, and $P=2p$ if $p$ is odd.
The first sum is running only over odd integers because the spin order is antiferromagnetic in $y$-direction, while the second sum is restricted to even integers since $\rho_j$ is translation invariant in $y$-direction.
The order parameters in Eq.~\eqref{eq: gaps} can be expanded accordingly
\begin{subequations} \label{eq: gaps Fourier}
\begin{align}
 \Delta_j^s &=
 \sum_{n \, \text{odd}} \, \Delta_n^s \, e^{i \bQ_n \cdot \mathbf{r}_j} \, , \\
 \Delta_j^c &=
 \sum_{n \, \text{even}} \Delta_n^c \, e^{i \bQ_n \cdot \mathbf{r}_j} \, ,
\end{align}
\end{subequations}
with $\Delta_n^c = \frac{1}{2} {\bar U} \varrho_n$ and $\Delta_n^s = {\bar U} M_n$.
Since $\Delta_j^c$ and $\Delta_j^s$ are real numbers, their Fourier components obey the relations $\Delta_n^c = (\Delta_{P-n}^c)^*$ and $\Delta_n^s = (\Delta_{P-n}^s)^*$.

Inserting Eq.~\eqref{eq: gaps Fourier} into Eq.~\eqref{eq: MF decoupling}, and Fourier transforming, yields the mean-field Hamiltonian in momentum representation,
\begin{equation} \label{eq: H_MF momentumspace}
\begin{split}
 H_\mathrm{MF} & = \int_\bk \epsilon_\bk\, \psi^\dag_\bk \psi_\bk \\
 & + \int_\bk \sum_{n=0}^{P-1} \left(
 \Delta_n^c \, \psi^\dag_\bk \sg^0 \psi_{\bk+\bQ_n} -
 \Delta_n^s \, \psi^\dag_\bk \sg^1 \psi_{\bk+\bQ_n} \right) \\
 & - \sum_{n=0}^{P-1} \left(
 {\bar U}^{-1} |\Delta_n^c|^2 - {\bar U}^{-1} |\Delta_n^s|^2 \right) \, ,
\end{split}
\end{equation}
where $\Delta_n^c = 0$ for odd $n$ and $\Delta_n^s = 0$ for even $n$.
Absorbing $\Delta_0^c$ into the chemical potential we can also set $\Delta_0^c = 0$.

Introducing a "Nambu spinor" with $P$ components
\begin{equation} \label{eq: Nambu spinor}
 \Psi_{\bk,\sigma} = \left(
 \begin{array}{c}
 \psi_{\bk,\sg} \\
 \psi_{\bk+\bQ_1,\bar\sg} \\
 \psi_{\bk+\bQ_2,\sg} \\
 \vdots \\
 \psi_{\bk+\bQ_{P-2},\sg} \\
 \psi_{\bk+\bQ_{P-1},\bar\sg}
 \end{array} \right) ,
\end{equation}
with the convention $\bar\up = \, \down$ and $\bar\down = \, \up$, one can cast the Hamiltonian~\eqref{eq: H_MF momentumspace} in the form
\begin{equation} \label{eq: HMF explicit}
 H_\mathrm{MF} =
 \sum_\sigma \int_\bk' \Psi^\dagger_{\bk,\sigma} \mathcal{H}_{\bk}
 \Psi_{\bk,\sigma} ,
\end{equation}
where we have dropped the constant term in Eq.~\eqref{eq: H_MF momentumspace}.
The momentum integration $\int_\bk' = \int_{\bk\in\mathrm{BZ'}}\frac{d^2\bk}{(2\pi)^2}$, extends over a reduced Brillouin zone $\mathrm{BZ'}$ corresponding to the reduced translation invariance in the stripe ordered state \cite{Scholle2024}.
The matrix $\mathcal{H}_{\bk}$ has $P \times P$ elements of the form
\begin{equation} \label{eq: H_MF components}
 \left[ \mathcal{H}_{\bk} \right]_{ll'} =
 \left\{ \begin{array}{lll}
 \epsilon_{\bk+\bQ_l} & \text{if} & l=l' \\
 -\Delta_n^s & \text{if} &
 l'= (l+n)\,\mathrm{mod}\,P, \, \text{$n$ odd} \\
 \Delta_n^c & \text{if} &
 l'= (l+n) \,\mathrm{mod} \, P, \, \text{$n$ even}
 \end{array} \right. .
\end{equation}
For example, for $P=6$, we have
\begin{equation}
 \mathcal{H}_{\bk} =
 \left(
 \begin{array}{cccccc}
        \epsilon_\bk & -\Delta_1^s & \Delta_2^c & -\Delta_3^s
        & \Delta_4^c  & -\Delta_5^s \\
        -\Delta_5^s & \epsilon_{\bk,1} & -\Delta_1^s & \Delta_2^c
        & -\Delta_3^s & \Delta_4^c \\
        \Delta_4^c & -\Delta_5^s & \epsilon_{\bk,2} & -\Delta_1^s
        & \Delta_2^c & -\Delta_3^s \\
        -\Delta_3^s & \Delta_4^c & -\Delta_5^s & \epsilon_{\bk,3}
        & -\Delta_1^s & \Delta_2^c \\
        \Delta_2^c & -\Delta_3^s & \Delta_4^c & -\Delta_5^s
        & \epsilon_{\bk,4} & -\Delta_1^s \\
        -\Delta_1^s & \Delta_2^c & -\Delta_3^s & \Delta_4^c
        & -\Delta_5^s & \epsilon_{\bk,5}
 \end{array}
 \right) \, ,
\end{equation}
where $\epsilon_{\bk,l} = \epsilon_{\bk+\bQ_l}$.

The matrix $\cH_\bk$ can be diagonalized by a unitary transformation $\cU_\bk$. The  corresponding eigenvalues $E_\bk^\ell$ with $\ell = 1,\dots,P$ describe the quasi-particle bands of the stripe state.
The mean-field free energy is thus given by
\begin{equation} \label{eq: free energy}
\begin{split}
 F = &- 2T \int'_\bk \sum_{\ell=1}^P \ln\left( 1 + e^{-E_\bk^\ell/T} \right) \\
 &- \sum_{n=0}^{P-1} \left(
 {\bar U}^{-1} |\Delta_n^c|^2 - {\bar U}^{-1} |\Delta_n^s|^2 \right) + \mu n \, ,
\end{split}
\end{equation}
where the chemical potential $\mu$ is determined by adjusting the density
$n = 2\int'_\bk \sum_\ell f(E_\bk^\ell)$ to a given value.


\subsubsection{Chargon Green function} \label{sec: chargon G}

The chargon Green function is defined as the expectation value
\begin{equation}
 G_{\sg\sg'}(k,k') = - \bra \psi_\sg(k) \psi_{\sg'}^*(k') \ket \, .
\end{equation}
$G_{\sg\sg}(k,k')$ is non-zero only for $k_0 = k'_0$ and $\bk = \bk' + \bQ_n$ with even $n$, while $G_{\sg\bar\sg}(k,k')$ is non-zero only for $k_0 = k'_0$ and $\bk = \bk' + \bQ_n$ with odd $n$.
Using the Grassmann field $\Psi_\sg(k)$ corresponding to the Nambu spinor $\Psi_{\bk\sg}$ defined in Eq.~\eqref{eq: Nambu spinor}, the non-zero contributions to $G_{\sg\sg'}(k,k')$ can be conveniently collected in a matrix Green function
\begin{equation} \label{eq: Gsg}
 \cG_\sg(k) = - \bra \Psi_\sg(k) \Psi_\sg^*(k) \ket \, =
 \left( ik_0 - \cH_\bk \right)^{-1} \, .
\end{equation}
Note that $\cG_\up = \cG_\down$ since $\cH_\bk$ is independent of $\sg$.
This is equivalent to the relations $G_{\up\up}(k,k') = G_{\down\down}(k,k')$ and $G_{\up\down}(k,k') = G_{\down\up}(k,k')$.
From $\cU_\bk^\dag \cH_\bk \cU_\bk = {\rm diag}\left( E_\bk^1,\dots,E_\bk^P \right)$ we obtain
$\cU_\bk^\dag \cG_\sg(k) \cU_\bk =
{\rm diag} \left[ (ik_0 - E_\bk^1)^{-1},\dots,(ik_0 - E_\bk^P)^{-1} \right]$, and thus
\begin{equation} \label{eq: eigenvalue decomp}
 \cG_\sg(k) = \sum_\ell \frac{g_\bk^\ell}{ik_0 - E_\bk^\ell} \, ,
\end{equation}
where $g_\bk^\ell$ is the $P \times P$ matrix formed from the elements
$(g_\bk^\ell)_{ll'} = (\cU_\bk)_{l\ell} (\cU_\bk^\dag)_{\ell l'}$.

Using the definition of the charge and magnetic gaps Eq.~\eqref{eq: gaps}, and their Fourier expansion Eq.~\eqref{eq: gaps Fourier}, one can write their self-consistency equations in terms of the Green function as
\begin{subequations} \label{eq: gap equations}
\begin{align}
 \Delta_n^c &= - {\bar U} \int_k \bra \psi_\sg(k+Q_n) \psi_\sg^*(k) \ket =
 {\bar U} \int_k G_{\sg\sg}(k+Q_n,k) \, , \\
 \Delta_n^s &= - {\bar U} \int_k \bra \psi_{\bar\sg}(k+Q_n) \psi_\sg^*(k) \ket =
 {\bar U} \int_k G_{\bar\sg\sg}(k+Q_n,k) \, ,
\end{align}
\end{subequations}
where $Q_n = (0,\bQ_n)$.
Using the spectral decomposition Eq.~\eqref{eq: eigenvalue decomp}, the Matsubara summation on  the right hand sides of these gap equations can be performed analytically.


\subsubsection{Chargon susceptibility} \label{sec: chargon chi}

The dynamical spin susceptibility of the chargons is defined as the connected expectation value
\begin{equation}
 \chi^{ab}(x,x') = \bra S_\psi^a(x) S_\psi^b(x') \ket_c \, ,
\end{equation}
where the variable $x = (\br,-i\tau)$ comprises the imaginary time $\tau$ and the lattice vectors $\br = \br_j$, while $S_\psi^a(x) = \frac{1}{2} \psi^*(x) \sg^a \psi(x)$ is the chargon pseudospin field expressed as a bilinear form of the chargon fields.


\subsection{Ward identity}

In this section we derive a Ward identity relating the gauge kernel $K_{\mu\nu}^{ab}$ to the spin susceptibility $\chi^{ab}$ in the stripe ordered state. This identity shows that the spin stiffness obtained from the gauge kernel is equivalent to a stiffness defined via the spin susceptibility. Similar Ward identities have already been obtained for N\'eel and spiral order \cite{Bonetti2022ward,Bonetti2024,Goremykin2024}.
In this section, we use a summation convention also for repeated spin indices.

The Ward identity involves a static symmetry breaking external field $\bh(\br)$ coupled linearly to the spin field $\bS(x)$. We choose this field to have the same spatial form as the expectation value of $\bS(x)$ in the ordered state,
\begin{equation}
 \bh(\br) = \frac{h}{m} \, {\bf m}(\br) = \frac{h}{m} \, \bra \bS(x) \ket \, ,
\end{equation}
where $h$ parametrizes the amplitude of the external field, and $m$ is the amplitude of the order parameter defined as
\begin{equation} \label{eq: m}
 m^2 = L^{-1} \sum_\br [{\bf m}(\br)]^2 = P^{-1} \sum_{\bar\br} [{\bf m}(\bar\br)]^2 \, ,
\end{equation}
where $L$ is the number of lattice sites, and the sum over $\bar\br$ is restricted to one unit cell (containing $P$ lattice sites).
Note that $\bra \bS(x) \ket$ does not depend on time.

From global SU(2) symmetry one can derive the following identity \cite{Bonetti2024}
\begin{equation}
 \int_{x,x'} \eps^{abc} \eps^{a'b'c'} \chi^{bb'}(x,x') \, h^c(\br) h^{c'}(\br') =
 \frac{m}{h} \int_x (\delta_{aa'} \delta_{cc'} - \delta_{ac} \delta_{a'c'}) \,
 h^c(\br) h^{c'}(\br) \, ,
\end{equation}
which is valid for any magnetic order. Specifying stripe order with all spins (and thus $h^a(\br)$) aligned in $x$-direction, this relation simplifies to
\begin{equation} \label{eq: global SU(2)}
 \int_{x,x'} \chi^{aa'}(x,x') \, h^1(\br) h^1(\br') =
 \left\{ \begin{array}{lll}
 {\displaystyle \frac{m}{h} \int_x [h^1(\br)]^2} &
 \mbox{for} & a = a' \in \{2,3\} \\
 0 & \mbox{else} &
 \end{array} \right. \, .
\end{equation}
We define a {\em weighted}\/ spin correlation function as
\begin{equation} \label{eq: def chiSS}
 \chit^{aa'}(x,x') =
 \frac{1}{m^2} m^1(\br) m^1(\br') \, \chi^{aa'}(x,x') \, ,
\end{equation}
and denote its Fourier transform by $\chit^{ab}(q,q')$.
Substituting $h^1(\br)$ by $S^1(\br)$ and using Eq.~\eqref{eq: m}, Eq.~\eqref{eq: global SU(2)} can then be expressed as
\begin{equation} \label{eq: chiSS global}
 \left. \chit^{aa'}(q,q') \right|_{q=q'=0} =
 \left\{ \begin{array}{lll}
 m/h & \mbox{for} & a = a' \in \{2,3\} \\
 0 & \mbox{else} &
 \end{array} \right. \, .
\end{equation}

From local SU(2) symmetry one can derive the Ward identity \cite{Bonetti2024}
\begin{eqnarray} \label{eq: wardid1a}
 \partial_\mu \partial'_\nu K_{\mu\nu}^{aa'}(x,x') &=&
 - \eps^{abc} \eps^{a'b'c'} \chi^{bb'}(x,x') h^c(\br) h^{c'}(\br')  \nonumber \\
 &-& (2\delta_{ac} \delta_{a'c'} - \delta_{ac'} \delta_{a'c} -
 \delta_{aa'} \delta_{cc'})
 m^c(\br) h^{c'}(\br) \delta(x-x') \, .
\end{eqnarray}
The continuum limit for the spatial derivatives is legitimate here because we will evaluate the Ward identity in Fourier space for small momenta only.
For stripe order with all spins aligned in $x$-direction, this yields
\begin{equation} \label{eq: wardid1b}
 \partial_\mu \partial'_\nu K_{\mu\nu}^{aa'}(x,x') =
 \left\{ \begin{array}{lll}
 - \chi^{\bar a \bar a}(x,x') h^1(\br) h^1(\br') + m^1(\br) h^1(\br) \delta(x-x') &
 \mbox{for} & a = a' \in \{2,3\} \\
 \chi^{\bar a \bar a'}(x,x') h^1(\br) h^1(\br') &
 \mbox{for} & a \neq a' \in \{2,3\} \\
 0 & \mbox{else} &
 \end{array} \right. \, ,
\end{equation}
where $\bar 2 = 3$ and $\bar 3 = 2$.
Substituting $h^1(\br)$ by $m^1(\br)$ and Fourier transforming, one obtains
\begin{equation} \label{eq: wardid1c}
 (\bq,iq_0)_\mu (\bq',iq'_0)_\nu K_{\mu\nu}^{aa'}(q,q') =
 \left\{ \begin{array}{lll}
 - h^2 \chit^{\bar a \bar a}(q,q') + hm &
 \mbox{for} & a = a' \in \{2,3\} \\
 h^2 \chit^{\bar a \bar a'}(q,q') &
 \mbox{for} & a \neq a' \in \{2,3\} \\
 0 & \mbox{else} &
 \end{array} \right. \, .
\end{equation}
Eq.~\eqref{eq: chiSS global} then implies
\begin{equation}
 \left. (\bq,iq_0)_\mu (\bq',iq'_0)_\nu K_{\mu\nu}^{aa'}(q,q') \right|_{q=q'=0} = 0 \, ,
\end{equation}
for all $a,a'$.

Taking second derivatives with respect to momenta and frequencies, the Ward identity Eq.~\eqref{eq: wardid1c} relates the spatial and temporal stiffnesses defined in Eqs.~\eqref{eq: spatial stiffness} and \eqref{eq: temporal stiffness}, repectively, to the weighted susceptibility $\chit^{aa}$. In particular, $J_{\mu\nu}^{a1} = J_{\mu\nu}^{1a} = 0$ for all $a$, while
\begin{eqnarray}
 \label{eq: wardstiff1a}
 J_{\alf\beta}^{aa} &=&
 - \frac{1}{2} \lim_{h \to 0} \left. h^2 \partial_{q_\alf} \partial_{q_\beta}
 \chit^{\bar a \bar a}(q,q) \right|_{q=0} \, , \\
 \label{eq: wardstiff1b}
 Z^{aa} &=&
 - \frac{1}{2} \lim_{h \to 0} \left. h^2 \partial_{q_0}^2
 \chit^{\bar a \bar a}(q,q) \right|_{q=0} \, ,
\end{eqnarray}
for $a \in \{2,3\}$ and $\alf,\beta \in \{x,y\}$.
In the special case of N\'eel order one has $m^1(\br) = m e^{i\bQ\br}$ with $\bQ = (\pi,\pi)$, so that $\chit^{ab}(q,q) = \chi^{ab}(Q+q,Q+q)$ with $Q = (\bQ,0)$. The relations \eqref{eq: wardstiff1a} and \eqref{eq: wardstiff1b} then agree with those derived in Refs.~\cite{Bonetti2024, Goremykin2024}.
It is crucial to take the limit $h \to 0$ {\em after}\/ the limit $q \to 0$. Taking the limits in the opposite order would yield $J_{\mu\nu}^{aa'} = 0$ for all $a,a'$.

The equivalence of the $y$ and $z$ directions for a stripe state with all spins aligned in $x$-direction implies that $\chi^{22} = \chi^{33} \equiv \chi^\perp$,
so that $J_{\alpha\beta}^{22} = J_{\alpha\beta}^{33} \equiv J_{\alpha\beta}$
and $Z^{22} = Z^{33} \equiv Z$.
Eqs.~\eqref{eq: chiSS global}, \eqref{eq: wardstiff1a}, and \eqref{eq: wardstiff1b} are consistent with
\begin{equation}
 \chit^\perp(q,q) \sim
 \frac{m^2}{Z q_0^2 + J_{\alf\beta} q_\alf q_\beta + hm}
\end{equation}
for small $q$.

In Appendix \ref{app: alternative WI} we derive another Ward identity for a {\em modified}\/ gauge kernel $\bar K_{\mu\nu}^{ab}$, which is related to $K_{\mu\nu}^{ab}$ by a Legendre transform.


\subsection{Spinon stiffnesses}

We now derive expressions for the spin stiffnesses from the general formulae \eqref{eq: spatial stiffness} and \eqref{eq: temporal stiffness} in the case where the chargons are stripe ordered. The gauge kernel $K_{\mu\nu}^{ab}$ is determined by two-point and four-point expectation values of chargon operators via Eq.~\eqref{eq: gauge kernel}. We compute the diamagnetic contribution in mean-field theory, and the paramagnetic contribution in random phase approximation (RPA), which is the conserving approximation for four point functions consistent with mean-field theory for the free energy and the Green function \cite{Baym1961}.
The RPA consists of a plain mean-field contribution and an interaction correction.
\begin{figure}[tb]
\centering
\includegraphics[width=14cm]{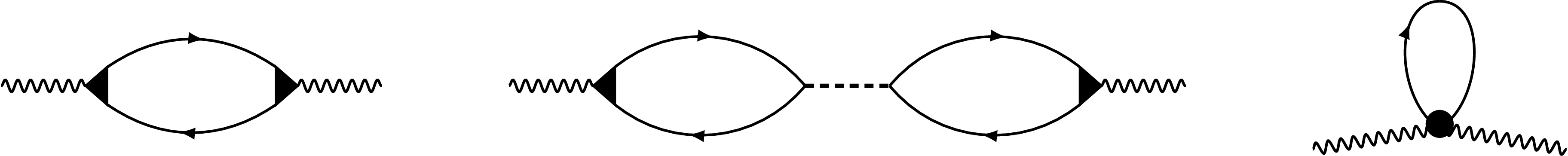}
\caption{Feynman diagrams representing the RPA contributions to the gauge kernel: bare paramagnetic contribution, interaction correction, diamagnetic contribution.}
\label{fig:RPA}
\end{figure}
The former can be represented diagrammatically by a bubble diagram, the latter by a pair of bubble diagrams connected by an effective interaction, see Fig.~1.


\subsubsection{Bare contributions} \label{sec: bare contributions}

The ``bare'' paramagnetic gauge kernel, associated with the bubble diagram, has the form
\begin{equation} \label{eq: Kp0}
 K_{0,\mu\nu}^{p,ab}(q) = \frac{1}{4} \int_{k,k'} \tr \big[
 \gam_{\bk+\bq}^{\mu,a} G(k+q,k'+q) \gam_{\bk'}^{\nu,b} G(k',k) \big] \, ,
\end{equation}
where $\gam_\bk^{\mu,a} = \gam_\bk^\mu \sg^a$, and
\begin{equation}
 G(k,k') = \left( \begin{array}{cc}
 G_{\up\up}(k,k') & G_{\up\down}(k,k') \\ G_{\down\up}(k,k') & G_{\down\down}(k,k')
 \end{array} \right) \, .
\end{equation}
To evaluate Eq.~\eqref{eq: Kp0} in a stripe state, we express the vertices and the Green functions in the spin-momentum basis ${\cal B} = ({\cal B}_\up,{\cal B}_\down)$ with
\begin{equation} \label{eq: basis B}
\begin{split}
 {\cal B}_\up &=
 ( \bk\!\up,\bk+\bQ_1\!\down,\bk+\bQ_2\!\up,\dots,\bk+\bQ_{P-1}\!\down ) \, , \\
 {\cal B}_\down &=
 ( \bk\!\down,\bk+\bQ_1\!\up,\bk+\bQ_2\!\down,\dots,\bk+\bQ_{P-1}\!\up ) \, ,
\end{split}
\end{equation}
where the momentum $\bk$ is restricted to the reduced Brillouin zone of the stripe state.
In this basis, the vertices are given by the $2P \times 2P$ matrices
\begin{subequations} \label{eq: Gammak}
\begin{align}
 \Gam_\bk^{\mu,1} &= \left( \begin{array}{cc}
 0 & \Gam_\bk^{\mu,+} \\ \Gam_\bk^{\mu,+} & 0  \end{array} \right) \, , \\[2mm]
 \Gam_\bk^{\mu,2} &= \left( \begin{array}{cc}
 0 & -i \Gam_\bk^{\mu,-} \\ i \Gam_\bk^{\mu,-} & 0  \end{array} \right) \, , \\[2mm]
 \Gam_\bk^{\mu,3} &= \left( \begin{array}{cc}
 \Gam_\bk^{\mu,-} & 0 \\ 0 & - \Gam_\bk^{\mu,-}  \end{array} \right) \, .
\end{align}
\end{subequations}
with the $P \times P$ matrices
\begin{equation}
 \Gam_\bk^{\mu,\pm} = {\rm diag} \left(
 \gam_\bk^\mu,\pm\gam_{\bk+\bQ_1}^\mu,\gam_{\bk+\bQ_2}^{\mu},\dots,
 \pm\gam_{\bk+\bQ_{P-1}}^\mu \right) \, .
\end{equation}
The Green function also becomes a $2P\times2P$ matrix in the basis $\cal B$,
\begin{equation} \label{eq: cG}
 \cG(k) = {\rm diag}[\cG_\up(k),\cG_\down(k)] \, ,
\end{equation}
where $\cG_\up(k)$ and $\cG_\down(k)$ are the (identical) $P\times P$ matrix Green functions defined in Eq.~\eqref{eq: Gsg}.

The bare paramagnetic gauge kernel in Eq.~\eqref{eq: Kp0} can then be written as
\begin{equation} \label{eq: Kp0 a}
 K_{0,\mu\nu}^{p,ab}(q) = \frac{1}{4} T \sum_{k_0} \int'_\bk \tr \big[
 \Gam_{\bk+\bq}^{\mu,a} \, \cG(k+q) \, \Gam_\bk^{\nu,b} \, \cG(k) \big] \, ,
\end{equation}
where the momentum integral extends over the reduced Brillouin zone.
Inserting Eqs.~\eqref{eq: Gammak} and \eqref{eq: cG} with the eigenvalue decomposition Eq.~\eqref{eq: eigenvalue decomp}, and performing the Matsubara sum, one obtains
\begin{eqnarray}
 \label{eq: K11}
 K_{0,\mu\nu}^{p,11}(q) &=&
 \frac{1}{2} \int'_\bk \sum_{\ell,\ell'} \tr\big(
 \Gam_{\bk+\bq}^{\mu,+} g_{\bk+\bq}^\ell \Gam_{\bk}^{\nu,+} g_\bk^{\ell'} \big)
 \frac{f(E_{\bk+\bq}^\ell) - f(E_\bk^{\ell'})}
 {iq_0 + E_{\bk+\bq}^\ell - E_\bk^{\ell'}} \, , \\
 \label{eq: K22}
 K_{0,\mu\nu}^{p,22}(q) &=& K_{0,\mu\nu}^{p,33}(q) \nonumber \\
 &=& \frac{1}{2} \int'_\bk \sum_{\ell,\ell'} \tr\big(
 \Gam_{\bk+\bq}^{\mu,-} g_{\bk+\bq}^\ell \Gam_{\bk}^{\nu,-} g_\bk^{\ell'} \big)
 \frac{f(E_{\bk+\bq}^\ell) - f(E_\bk^{\ell'})}
 {iq_0 + E_{\bk+\bq}^\ell - E_\bk^{\ell'}} \, ,
\end{eqnarray}
while all the off-diagonal components $K_{0,\mu\nu}^{p,ab}(q)$ with $a \neq b$ vanish.

The diamagnetic contribution to the gauge kernel is given by
\begin{equation} \label{eq: Kd1}
 K_{\alf\beta}^{d,ab} = \frac{1}{4} \delta_{ab} \int_k
 \gam_{\bk}^{\alf\beta} \tr[G(k,k)] \, ,
\end{equation}
where $G$ is the chargon Green function computed in mean-field theory.
It has only spatial components, that is, $\alf,\beta \in \{x,y\}$.
In the stripe order basis $\cal B$, the diamagnetic vertex has the $2P\times2P$ matrix structure
\begin{equation}
 \Gam_\bk^{d,\alf\beta} = \left( \begin{array}{cc}
 \Gam_\bk^{\alf\beta} & 0 \\ 0 & \Gam_\bk^{\alf\beta}  \end{array} \right) \, ,
\end{equation}
with the $P \times P$ matrix
\begin{equation}
 \Gam_\bk^{\alf\beta} = {\rm diag} \left(
 \gam_\bk^{\alf\beta},\gam_{\bk+\bQ_1}^{\alf\beta},\gam_{\bk+\bQ_2}^{\alf\beta},
 \dots,\gam_{\bk+\bQ_{P-1}}^{\alf\beta} \right) \, .
\end{equation}
The diamagnetic gauge kernel can thus be written as
\begin{equation} \label{eq: Kd2}
 K_{\alf\beta}^{d,ab} = \frac{1}{4} \delta_{ab} \, T \sum_{k_0} \int'_\bk
 \tr \left[ \Gam_\bk^{\alf\beta} \cG(k) \right] \, .
\end{equation}
Inserting Eq.~\eqref{eq: cG} with the eigenvalue decomposition Eq.~\eqref{eq: eigenvalue decomp}, and performing the Matsubara sum, one obtains
\begin{equation} \label{eq: Kd3}
 K_{\alf\beta}^{d,ab} = \frac{1}{2} \delta_{ab} \int'_\bk \sum_\ell
 \tr \left( \Gam_\bk^{\alf\beta} g_\bk^\ell \right) f(E_\bk^\ell) \, .
\end{equation}

We now compute the spatial stiffnesses by taking the limits $\bq \to 0$ and $q_0 \to 0$ as in Eq.~\eqref{eq: spatial stiffness}.
For the bare paramagnetic contribution the symmetry breaking external field $h$ is irrelevant and can be set equal to zero from the beginning. The diamagnetic contribution is a momentum and frequency independent constant.
The bare paramagnetic contribution to the spatial stiffnesses is obtained as
\begin{eqnarray}
 J_{0,\alf\beta}^{p,11} &=&
 \lim_{\bq \to \mathbf{0}} K_{0,\alf\beta}^{p,11}(\bq,0)  \nonumber \\
 &=& \frac{1}{2} \int'_\bk \sum_{\ell,\ell'} \tr\big(
 \Gam_{\bk}^{\alf,+} g_{\bk}^\ell \Gam_{\bk}^{\beta,+} g_\bk^{\ell'} \big)
 \left[ \delta_{\ell\ell'} f'(E_\bk^\ell) +
 (1-\delta_{\ell\ell'}) \frac{f(E_{\bk}^\ell) - f(E_\bk^{\ell'})}
 {E_{\bk}^\ell - E_\bk^{\ell'}} \right] \, , \\
 J_{0,\alf\beta}^{p,22} &=& J_{0,\alf\beta}^{p,33} =
 \lim_{\bq \to \mathbf{0}} K_{0,\alf\beta}^{p,22}(\bq,0) =
 \lim_{\bq \to \mathbf{0}} K_{0,\alf\beta}^{p,33}(\bq,0) \nonumber \\
 &=& \frac{1}{2} \int'_\bk \sum_{\ell,\ell'} \tr\big(
 \Gam_{\bk}^{\alf,-} g_{\bk}^\ell \Gam_{\bk}^{\beta,-} g_\bk^{\ell'} \big)
 \left[ \delta_{\ell\ell'} f'(E_\bk^\ell) +
 (1-\delta_{\ell\ell'}) \frac{f(E_{\bk}^\ell) - f(E_\bk^{\ell'})}
 {E_{\bk}^\ell - E_\bk^{\ell'}} \right] \, .
\end{eqnarray}
In Appendix \ref{app: cancellation} we show that
$J_{0,\alf\beta}^{p,11} = - K_{\alf\beta}^{d,11}$,
so that the diamagnetic and the bare paramagnetic contributions to $J_{\alf\beta}^{11}$ cancel each other.

The temporal stiffness $Z^{ab}$ defined in Eq.~\eqref{eq: temporal stiffness} has only paramagnetic contributions. The bare contributions are obtained from Eqs.~\eqref{eq: K11} and \eqref{eq: K22} as
\begin{eqnarray}
 \label{eq: Z11}
 Z_0^{11} &=&
 - \frac{1}{2} \int'_\bk \sum_{\ell\neq\ell'} \tr\big(
 \Gam^{0,+} g_{\bk}^\ell \Gam^{0,+} g_\bk^{\ell'} \big)
 \frac{f(E_{\bk}^\ell) - f(E_\bk^{\ell'})}
 {E_{\bk}^\ell - E_\bk^{\ell'}} = 0 \, , \\
 \label{eq: Z22}
 Z_0^{22} &=&
 - \frac{1}{2} \int'_\bk \sum_{\ell\neq\ell'} \tr\big(
 \Gam^{0,-} g_{\bk}^\ell \Gam^{0,-} g_\bk^{\ell'} \big)
 \frac{f(E_{\bk}^\ell) - f(E_\bk^{\ell'})}
 {E_{\bk}^\ell - E_\bk^{\ell'}} = Z_0^{33} \, .
\end{eqnarray}
The vertices $\Gam^{0,\pm}$ are momentum independent.
The first component $Z_0^{11}$ vanishes because $\Gam^{0,+}$ is the identity matrix and $\tr(g_{\bk}^\ell g_\bk^{\ell'}) = 0$ for $\ell \neq \ell'$.


\subsubsection{Interaction correction} \label{sec: interaction correction}

The interaction correction to $K_{\mu\nu}^{ab}(q)$ has the form
\begin{equation} \label{eq: Delta K}
 \Delta K_{\mu\nu}^{ab}(q) = \int_{q',q''} \sum_{c,d}
 K_{0,\mu0}^{p,ac}(q,q') \Gam^{cd}(q',q'') K_{0,0\nu}^{p,db}(q'',q) \, ,
\end{equation}
where $K_{0,\mu\nu}^{p,ab}(q,q')$ is the bare paramagnetic gauge kernel with arbitrary ingoing and outgoing momenta, and $\Gam$ is the RPA effective interaction. The latter is determined by the linear integral equation
\begin{equation} \label{eq: eff interaction}
 \Gam^{cd}(q,q') = \Gam_0^{cd} + \int_{q''} \sum_{c'd'}
 \Gam_0^{cc'} \chi_0^{c'd'}(q,q'') \Gam^{d'd}(q'',q') \, ,
\end{equation}
with the bare interaction $\Gam_0 = 2 \, {\rm diag}(-{\bar U},{\bar U},{\bar U},{\bar U})$, and the bare susceptibility
$\chi_0^{cd}(q,q') = - K_{0,00}^{cd}(q,q') = - K_{0,00}^{p,cd}(q,q')$.
While the calculation of the stiffnesses requires only the spin SU(2) gauge kernel $K_{\mu\nu}^{ab}(q)$ with indices $a,b = 1,2,3$, the effective interaction couples spin and charge, so that the indices $c$ and $d$ in Eqs.~\eqref{eq: Delta K} and \eqref{eq: eff interaction} are spin-charge indices, and the Pauli matrices in the spin channels are complemented by the unit matrix $\sg^0$ in the charge channel.

The above equations involve off-diagonal components of the bare gauge kernels, both in the momentum and spin variables. They are still given by bubble diagrams, but now the ingoing and outgoing momenta may differ by $\bQ_n = n(2\pi/p,\pi)$ with $n = 0,\dots,P-1$.
Generalizing Eq.~\eqref{eq: Kp0}, we obtain
\begin{equation} \label{eq: Kp0'}
 K_{0,\mu\nu}^{p,ab}(q,q') = \frac{1}{4} \int_{k,k'} \tr \big[
 \gam_{\bk+\bq}^{\mu,a} G(k+q,k'+q') \gam_{\bk'}^{\nu,b} G(k',k) \big] \, ,
\end{equation}
where $\gam_\bk^{\mu,a} = \gam_\bk^\mu \sg^a$ now includes also the charge vertex for $a=0$.

To solve the integral equation \eqref{eq: eff interaction}, and to evaluate
$\Delta K_{\mu\nu}^{ab}(q)$, it is convenient to cast the off-diagonal momentum dependences in a matrix form. To this end, we substitute $\bq \to \bq + \bQ_n$, $\bq' \to \bq' + \bQ_{n'}$, etc., and restrict the continuous part of the momenta to the magnetic Brillouin zone.
Since $K_{0,\mu\nu}^{p,ab}(q,q')$ and $\Gam^{cd}(q,q')$ are diagonal in the frequency variables $q_0$ and $q'_0$, and also in the restricted momentum variables $\bq$ and $\bq'$, we can substitute $K_{0,\mu\nu}^{p,ab}(q,q') \to K_{0,\mu\nu}^{p,ab}(\bQ_n,\bQ_{n'};q)$ and
$\Gam^{cd}(q,q') \to \Gam^{cd}(\bQ_n,\bQ_{n'};q)$.
Eq.~\eqref{eq: Delta K} can then be written as
\begin{equation} \label{eq: Delta K a}
 \Delta K_{\mu\nu}^{ab}(q) = \sum_{c,d} \sum_{n',n''}
 K_{0,\mu0}^{p,ac}(\mathbf{0},\bQ_{n'};q) \Gam^{cd}(\bQ_{n'},\bQ_{n''};q)
 K_{0,0\nu}^{p,db}(\bQ_{n''},\mathbf{0};q) \, ,
\end{equation}
and Eq.~\eqref{eq: eff interaction} becomes
\begin{equation} \label{eq: eff interaction a}
 \Gam^{cd}(\bQ_n,\bQ_{n'};q) = \Gam_0^{cd} + \sum_{n''} \sum_{c'd'}
 \Gam_0^{cc'} \chi_0^{c'd'}(\bQ_n,\bQ_{n''};q) \Gam^{d'd}(\bQ_{n''},\bQ_{n'};q) \, .
\end{equation}
Note that the external momentum $\bq$ in Eq.~\eqref{eq: Delta K} can be assumed to lie in the magnetic Brillouin zone, since we are ultimately interested in the limit $q \to 0$.
Defining $4P \times 4P$ matrices $\Gam(q)$ and $\chi_0(q)$ with matrix elements
$\Gam_{nn'}^{cd}(q) = \Gam^{cd}(\bQ_n,\bQ_{n'};q)$ and $\chi_0^{cd}(\bQ_n,\bQ_{n'};q)$, respectively, where $n,c$ are row indices and $n',d$ column indices, Eq.~\eqref{eq: eff interaction a} can be formally solved as
\begin{equation} \label{eq: eff interaction b}
 \Gam(q) = \left[ 1 - \Gam_0 \chi_0(q) \right]^{-1} \Gam_0 \, ,
\end{equation}
with $\Gam_{0;nn'} = 2 \, {\rm diag}(-{\bar U},{\bar U},{\bar U},{\bar U}) \, \delta_{nn'}$.

Using the matrix notation introduced in Sec.~\ref{sec: bare contributions}, the bare paramagnetic gauge kernel in Eq.~\eqref{eq: Kp0'} can be written as
\begin{equation} \label{eq: Kp0' a}
 K_{0,\mu\nu}^{p,ab}(\bQ_n,\bQ_{n'};q) = \frac{1}{4} \int'_k \tr
 \left[ \Gam_{\bk+\bq+\bQ_n}^{\mu,a} \Pi_n \cG(k+q) \Pi_{n'}^T
 \Gam_\bk^{\nu,b} \cG(k) \right] \, ,
\end{equation}
where $\Pi_n$ denotes a cyclic permutation by $n$ in the bases $\cB_\up$ and $\cB_\down$, so that $\Pi_n \cG \Pi_{n'}^T$ denotes the matrix obtained from $\cG = {\rm diag}(\cG_\up,\cG_\down)$ by shifting the rows of each $\cG_\sg$ by $n$ and the columns by $n'$. This formula generalizes Eq.~\eqref{eq: Kp0 a}, which holds only in the special case $\bQ_n = \bQ_{n'} = 0$.
In particular, the bare susceptibility $\chi_0^{ab} = - K_{0,00}^{p,ab}$ can be written as
\begin{equation}
 \chi_0^{ab}(\bQ_n,\bQ_{n'};q) = - \frac{1}{4} \int'_k \tr
 \left[ \Sg^a \Pi_n \cG(k+q) \Pi_{n'}^T \Sg^b \cG(k) \right] \, ,
\end{equation}
where $\Sg^0$ is the $2P \times 2P$ unit matrix, and $\Sg^a = \Gam^{0,a}$ with $a=1,2,3$ are block Pauli matrices in the basis $\cB$, see Eq.~\eqref{eq: Gammak}.
In Appendix~\ref{app: susc sym} we derive symmetry relations for
$\chi_0^{ab}(\bQ_n,\bQ_{n'};q)$, from which one obtains some zeros at $q=0$.

Using the eigenvalue decomposition \eqref{eq: eigenvalue decomp} of $\cG_\sg(k)$, the Matsubara sum over $k_0$ in Eq.~\eqref{eq: Kp0' a} can be carried out. Continuing the Matsubara frequency $q_0$ to real frequencies, $iq_0 \to \omega + i0^+$, we obtain
\begin{equation} \label{eq: Kp0' b}
 K_{0,\mu\nu}^{p,ab}(\bQ_n,\bQ_{n'};\bq,\omega) =
 \frac{1}{2} \int'_\bk \sum_{\ell,\ell'}
 A_{\mu\nu,\ell\ell'}^{ab}(\bQ_n,\bQ_{n'};\bk,\bq) F_{\ell\ell'}(\bk,\bq,\omega) \, ,
\end{equation}
with
\begin{equation} \label{eq: Amunu}
 A_{\mu\nu,\ell\ell'}^{ab}(\bQ_n,\bQ_{n'};\bk,\bq) =
 \frac{1}{2} \tr \left[ \Gam_{\bk+\bq+\bQ_n}^{\mu,a}
 \left( \begin{array}{cc} \Pi_n g_{\bk+\bq}^\ell \Pi_{n'}^T & 0 \\[-2mm]
 0 & \Pi_n g_{\bk+\bq}^\ell \Pi_{n'}^T \end{array} \right)
 \Gam_\bk^{\nu,b}
 \left( \begin{array}{cc} g_\bk^{\ell'} & 0 \\[-2mm]
 0 & g_\bk^{\ell'} \end{array} \right) \right] \, ,
\end{equation}
and
\begin{equation} \label{eq: Fll'}
 F_{\ell\ell'}(\bk,\bq,\omega) =
 \frac{f(E_\bk^\ell) - f(E_{\bk+\bq}^{\ell'})}
 {\omega + i0^+ + E_\bk^\ell - E_{\bk+\bq}^{\ell'}} \, .
\end{equation}

To compute the interaction correction to the stiffnesses, we need to take the limits $\bq \to \mathbf{0}$, $q_0 \to 0$, and $h \to 0$ in the appropriate order, see Eqs.~\eqref{eq: spatial stiffness} and \eqref{eq: temporal stiffness}. We recall that $h$ is a symmetry breaking field.
Although the effective interaction diverges for $\bq,q_0,h \to 0$ due to its Goldstone poles, the correction to the gauge kernel remains generally finite, thanks to compensating zeros in the bare gauge kernels, between which $\Gam$ is sandwiched in Eq.~\eqref{eq: Delta K}.
However, the limit value depends on the order in which $\bq$, $q_0$, and $h$ are sent to zero.
In the case of spiral order (including N\'eel order as a special case), it is possible to take the limits analytically, and derive manifestly finite interaction corrections to the stiffnesses \cite{Bonetti2022a, Bonetti2022ward, Bonetti2024}.
The terms which are removed by the symmetry breaking field $h$ have qualitatively the asymptotic form $\frac{q^2}{q^2+h}$, where $q$ stands for the small momentum or frequency variable. They vanish if $q \to 0$ before $h \to 0$, while giving a finite result in the opposite order of limits.
For the case of stripe order with a period $p > 2$ it seems difficult to isolate the singular terms by hand, and to derive analytic expressions for the stiffnesses. However, we can still isolate the singular terms numerically, by diagonalizing the matrix denominator $1 - \Gam_0 \chi_0(q)$ in Eq.~\eqref{eq: eff interaction b}. Some of the eigenvalues vanish for $\bq,q_0,h \to 0$. We then project out the corresponding eigenspaces, and perform the matrix inversion only in the remaining orthogonal space, yielding the socalled pseudoinverse \cite{Penrose1955}, which remains finite by construction. Inserting the resulting effective interaction into Eq.~\eqref{eq: Delta K a} simulates the limit defined with an external field $h$, without the need to actually calculate any quantity in the presence of $h$. The order of the limits $\bq \to \mathbf{0}$ and $q_0 \to 0$ still matters, and must be performed as described in the defining equations \eqref{eq: spatial stiffness} and \eqref{eq: temporal stiffness} for the spatial and temporal stiffnesses, respectively.

Like the bare contributions, the interaction corrections to the stiffnesses $\Delta J_{\mu\nu}^{11}$ and $\Delta J_{\mu\nu}^{ab}$ with $a \neq b$ vanish, while $\Delta J_{\mu\nu}^{22} = \Delta J_{\mu\nu}^{33}$.
Hence, the total stiffness
$J_{\mu\nu}^{ab} = J_{0,\mu\nu}^{p,ab} + J_{0,\mu\nu}^{d,ab} + \Delta J_{\mu\nu}^{ab}$ has the diagonal matrix form
\begin{equation}
 \cJ_{\mu\nu} = {\rm diag}(0,J_{\mu\nu},J_{\mu\nu}) \, .
\end{equation}
This simple form could have been obtained also directly from the symmetry breaking pattern
$\rm SU(2) \to U(1)$, where the residual U(1) symmetry of the broken SU(2) contains rotations around the axis of the collinear spin alignment (the $x$-axis in spin space in our convention).
For a stripe state with N\'eel order along one of the x or y (spatial) directions, the off-diagonal spatial stiffness components vanish, that is, $J_{xy} = J_{yx} = 0$. The diagonal components $J_{xx}$ and $J_{yy}$ are not equal, except in the special case of the N\'eel state. We neglect the mixed spatio-temporal components $J_{\alf0}$ and $J_{0\beta}$, which may occur due to Landau damping of the spin waves.


\subsection{Solution of nonlinear sigma model}

We solve the nonlinear sigma model \eqref{eq: nlsm} in a saddle-point approximation in the $\rm CP^1$ representation, which becomes exact in a large $N$ limit \cite{Auerbach1994}.


\subsubsection{\texorpdfstring{CP\/$^1$}{CP} representation}

The matrix $\cR$ can be expressed as a triad of orthonormal unit vectors
$\cR = (\hat\bn^1,\hat\bn^2,\hat\bn^3$), which can be represented in terms of two complex Schwinger bosons $z_\up$ and $z_\down$ \cite{Sachdev1995},
\begin{subequations} \label{eq: ni}
\begin{align}
 & {\bf\hat n}^1 = z^* \vec{\sigma} z , \\
 & {\bf\hat n}^- = z (i\sigma^2\vec{\sigma}) z , \\
 & {\bf\hat n}^+ = z^* (i\sigma^2\vec{\sigma})^\dag z^* ,
\end{align}
\end{subequations}
with $z = (z_\up,z_\down)$ and ${\bf\hat n}^\pm = {\bf\hat n}^2 \mp i {\bf\hat n}^3$.
The Schwinger bosons obey the constraint
\begin{equation} \label{eq: z boson constraint}
 z^*_\up z_\up + z^*_\down z_\down = 1 \, .
\end{equation}
The parametrization~\eqref{eq: ni} is equivalent to
\begin{equation} \label{eq: R to z}
 R = \left( \begin{array}{cc}
 z_\up &  -z_\down^* \\ z_\down & \phantom{-} z_\up^*
 \end{array} \right).
\end{equation}
Inserting this representation of $\cR$ into Eq.~\eqref{eq: nlsm} with a stiffness matrix of the form $\cJ_{\mu\nu} = {\rm diag}(0,J_{\mu\nu},J_{\mu\nu})$, we obtain the $\rm CP^1$ action
\begin{equation} \label{eq: CP1 action}
 \mathcal{S}_{\text{CP}^1}[z,z^*] = \int dx \,
 \Big[ 2J_{\mu\nu} [\partial_\mu z^*(x)] \partial_\nu z(x)
 - \, 2J_{\mu\nu} j_\mu(x) j_\nu(x) \Big] \, ,
\end{equation}
with the current density
\begin{equation} \label{eq: current}
 j_\mu(x) = \frac{i}{2}
 \left[ z^*(x) \partial_\mu z(x) - z(x) \partial_\mu z^*(x) \right] \, .
\end{equation}
%


\subsubsection{Large {\em N}\/ limit and saddle point}

The current-current interaction in Eq.~\eqref{eq: CP1 action} can be decoupled by a Hubbard-Stratonovich transformation with a U(1) gauge field $\mathcal{A}_\mu$, and the constraint \eqref{eq: z boson constraint} can be implemented by a Lagrange multiplier field $\lam$, leading to the action \cite{Auerbach1994}
\begin{equation} \label{eq: CP1 model}
 \cS_{\text{CP}^1}[z,z^*,\mathcal{A}_\mu,\lam] =
 \int dx \big[ 2J_{\mu\nu} (D_\mu z)^* (D_\nu z)
 + i\lambda(z^*z-1) \big] \, ,
\end{equation}
where $D_\mu = \partial_\mu - i\mathcal{A}_\mu$ is the covariant derivative. We have suppressed the space-time dependence of the fields to shorten the expression.

To perform a large $N$ expansion, we extend the two-component field $z = (z_\up,z_\down)$ to an $N$-component field $z = (z_1,\dots,z_N)$, and rescale it by a factor $\sqrt{N/2}$ so that it now satisfies the local constraint
\begin{equation}
 z^*(x) z(x) = \sum_{\sg=1}^N z^*_\sg(x) z_\sg(x) = \frac{N}{2} \, .
\end{equation}
To obtain a nontrivial limit $N \to \infty$, we rescale also the stiffnesses $J_{\mu\nu}$ by a factor $2/N$, yielding the CP$^{N-1}$ action
\begin{equation} \label{eq: CPN1 model}
 \cS_{\text{CP}^{N-1}}[z,z^*,\mathcal{A}_\mu,\lam] =
 \int dx \big[ 2J_{\mu\nu} (D_\mu z)^* (D_\nu z)
 + i\lam \left( z^*z - N/2 \right) \big] .
\end{equation}
In the large $N$ limit, the partition function of the CP$^{N-1}$ model is dominated by a saddle point for the fields $\mathcal{A}$ and $\lambda$, at which $\mathcal{A}_\mu(x) = 0$, and $\lam(x) = \lam$ is uniform in space and time \cite{Auerbach1994}.
One can then read off the spinon propagator directly from the action \eqref{eq: CPN1 model},
\begin{equation} \label{eq: spinon propag}
 D(q) = \bra z_\sg^*(q) z_\sg(q) \ket = \frac{1}{2Z (m_s^2 + q_0^2) +
 2J_{\alpha\beta} q_\alpha q_\beta} \, ,
\end{equation}
where $2m_s^2 = i\lam/Z$.
The Lagrange multiplier $\lam$ and thus the spinon mass $m_s$ are fixed by taking the average of the constraint, leading to $\bra z^*(x) z(x) \ket = N/2$.
At finite temperatures, and also for a quantum disordered ground state, there is no Bose condensate, that is $\bra z(x) \ket = 0$, and the constraint leads to the condition
\begin{equation} \label{eq: constraint}
 2 \int_q D(q) =
 \int_q \frac{1}{Z (m_s^2 + q_0^2) + J_{\alpha\beta} q_\alpha q_\beta} = 1 \, ,
\end{equation}
which determines the spinon mass $m_s$ as a function of $Z$, $J_{\alpha\beta}$, and an ultraviolet cutoff.
Choosing an isotropic momentum cutoff $\Lam_{\rm uv}$, and performing the Matsubara sum over $q_0$, the condition \eqref{eq: constraint} can be written more explicitly as
\begin{equation}
 \frac{1}{4\pi J} \int_0^{c_s \Lam_{\rm uv}}
 \frac{\eps \, d\eps}{\sqrt{m_s^2 + \eps^2}}
 \coth\left[\sqrt{m_s^2 + \eps^2}/(2T)\right] = 1 \, ,
\end{equation}
where
\begin{equation}
 J = \sqrt{{\rm det} \left( \begin{array}{ll}
 J_{xx} & J_{xy} \\[-2.5mm] J_{yx} & J_{yy} \end{array} \right)} \, ,
\end{equation}
is an ``average'' spin stiffness, and $c_s = \sqrt{J/Z}$ the corresponding average spin wave velocity. For stripe order with a N\'eel pattern along one of the axes (in $x$ or $y$ direction), the off-diagonal components of $J_{\alf\beta}$ vanish, so that
$J = \sqrt{J_{xx} J_{yy}}$.

The spinon propagator Eq.~\eqref{eq: spinon propag} can be written in the alternative form
\begin{equation}
 D(q) = - \frac{1}{2Z} \frac{1}{(iq_0 - \omega_\bq)(iq_0 + \omega_\bq)} \, ,
\end{equation}
with the spinon dispersion
\begin{equation}
\omega_\bq = \sqrt{m_s^2 + (J_{\alf\beta} \, \bq_\alf \bq_\beta)/Z} \, .
\end{equation}
From this expression it is evident that $D(q)$ describes propagating bosonic excitations with a gapped dispersion $\omega_\bq$. For $m_s \to 0$ one recovers the usual spin waves of an ordered magnet.


\subsection{Electron Green function}

We now derive an expression for the electron Green function $G^e$, from which the spectral function for single-particle excitations can be obtained. The superscript ``e'' distinguishes the electron Green function from the chargon Green function $G$.
The imaginary time electron Green function is defined as
\begin{equation}
 G_{j\sg,j'\sg'}^e(\tau) = - \bra c_{j\sg}(\tau) c^*_{j'\sg'}(0) \ket \, .
\end{equation}
Inserting the fractionalization \eqref{eq: electron fract} of the electron fields, we obtain
\begin{equation}
 G_{j\sg,j'\sg'}^e(\tau) =
 - \sum_{s,s'} \bra [R_j(\tau)]_{\sg s} [R_{j'}^*(0)]_{\sg's'}
 \psi_{js}(\tau) \psi_{j's'}^*(0) \ket \, .
\end{equation}
We simplify this expression by neglecting chargon-spinon interactions, so that we can decouple the expectation value as
\begin{equation} \label{eq: Ge factorized}
 G_{j\sg,j'\sg'}^e(\tau) =
 - \sum_{s,s'} \bra [R_j(\tau)]_{\sg s} [R_{j'}^*(0)]_{\sg's'} \ket
 \bra \psi_{js}(\tau) \psi_{j's'}^*(0) \ket \, .
\end{equation}
This approximation is common in gauge theories of fluctuating magnetic order \cite{Borejsza2004, Scheurer2018, Wu2018, Bonetti2022a}. The first expectation value is the spinon propagator, and the second one the chargon Green function in real space representation, $G_{js,j's'}(\tau) = - \bra \psi_{js}(\tau) \psi_{j's'}^*(0) \ket$.

Using the Schwinger boson representation \eqref{eq: R to z} of $R$, we obtain
\begin{equation}
 \bra [R_j(\tau)]_{\sg s} [R_{j'}^*(0)]_{\sg's'} \ket =
 \bra z_{j\sg}(\tau) z_{j'\sg}^*(0) \ket \delta_{\sg\sg'} \delta_{ss'} =
 D_{jj'}(\tau) \delta_{\sg\sg'} \delta_{ss'} \, ,
\end{equation}
in the saddle point approximation to the nonlinear sigma model, where $D_{jj'}(\tau)$ is the Fourier transform of the spinon propagator $D(q)$ in Eq.~\eqref{eq: spinon propag}.
The Kronecker deltas arise since expectation values of the form $\bra z_\sg z_{\sg'} \ket$ and $\bra z_\up^* z_\down \ket$ vanish due to the SU(2) symmetry of the nonlinear sigma model.
Fourier transforming Eq.~\eqref{eq: Ge factorized}, we then obtain the electron Green function in momentum representation, $G_{\sg\sg'}^e(\bk,\bk',ik_0) = G^e(\bk,\bk',ik_0) \, \delta_{\sg\sg'}$, where
\begin{equation} \label{eq: Ge mom rep}
 G^e(\bk,\bk',ik_0) =
 T \sum_{q_0} \int_\bq \sum_s G_{ss}(\bk-\bq,\bk'-\bq,ik_0-iq_0) \, D(\bq,iq_0) \, .
\end{equation}
At this point we see that the electron Green function is SU(2) symmetric, that is, the spinon fluctuations restore the broken SU(2) symmetry of the chargons. However, the translation symmetry remains partially broken, that is, $G^e(\bk,\bk',ik_0)$ has momentum off-diagonal contributions for $\bk'-\bk = \bQ_n$ with even $n$. This is due to the charge order of the stripes, which is not removed by the spinon fluctuations. By contrast, in a fluctuating spiral magnet there is no charge order, and the electron Green function recovers the full translation invariance of the lattice \cite{Bonetti2022a}.
Substituting $\bk \to \bk + \bQ_n$ and $\bk' \to \bk' + \bQ_{n'}$, where $\bk$ and $\bk'$ are situated in the first magnetic Brillouin zone, we have $\bk = \bk'$ and the momentum dependence of the electron Green function can be parametrized as $G^e(\bQ_n,\bQ_{n'};\bk,ik_0)$, with $n'-n$ even.

Using the eigenvalue decomposition Eq.~\eqref{eq: eigenvalue decomp} of the chargon Green function, the Matsubara sum in Eq.~\eqref{eq: Ge mom rep} can be carried out, and the analytic continuation $ik_0 \to \omega + i0^+$ can be performed explicitly. After some straightforward algebra we obtain the electron Green function in its final form,
\begin{equation} \label{eq: Ge final}
 G^e(\bQ_n,\bQ_{n'};\bk,\omega) =
 \int_\bq \sum_{\ell} \sum_{p=\pm} \frac{1}{2Z \omega_\bq} \,
 \frac{b(\omega_\bq) + f(pE_{\bk-\bq}^\ell)}
 {\omega + i0^+ - E_{\bk-\bq}^\ell + p\,\omega_\bq}
 (g_{\bk-\bq}^\ell)_{nn'} \, .
\end{equation}
The momentum integral over $\bq$ is restricted by the same ultraviolet cutoff as the momentum integral of the constraint Eq.~\eqref{eq: constraint}.

In the special case of N\'eel order, that is, for $P=2$, the electron Green function is translation invariant, and the contributing spectral weights have the form
$(g_\bk^\ell)_{00} = 1 + \ell h_\bk/e_\bk$ and
$(g_\bk^\ell)_{11} = 1 + \ell h_{\bk+\bQ}/e_{\bk+\bQ}$, where
$h_\bk = \frac{1}{2} (\eps_\bk - \eps_{\bk+\bQ})$ and $e_\bk = \sqrt{\Delta^2 + h_\bk^2}$.
The expression \eqref{eq: Ge final} then agrees with the corresponding expression for a spiral magnet in the N\'eel limit $\bQ \to (\pi,\pi)$ derived in Ref.~\cite{Bonetti2022a}.


\section{Results}

\begin{figure}[tb]
\centering
\includegraphics[width=8cm]{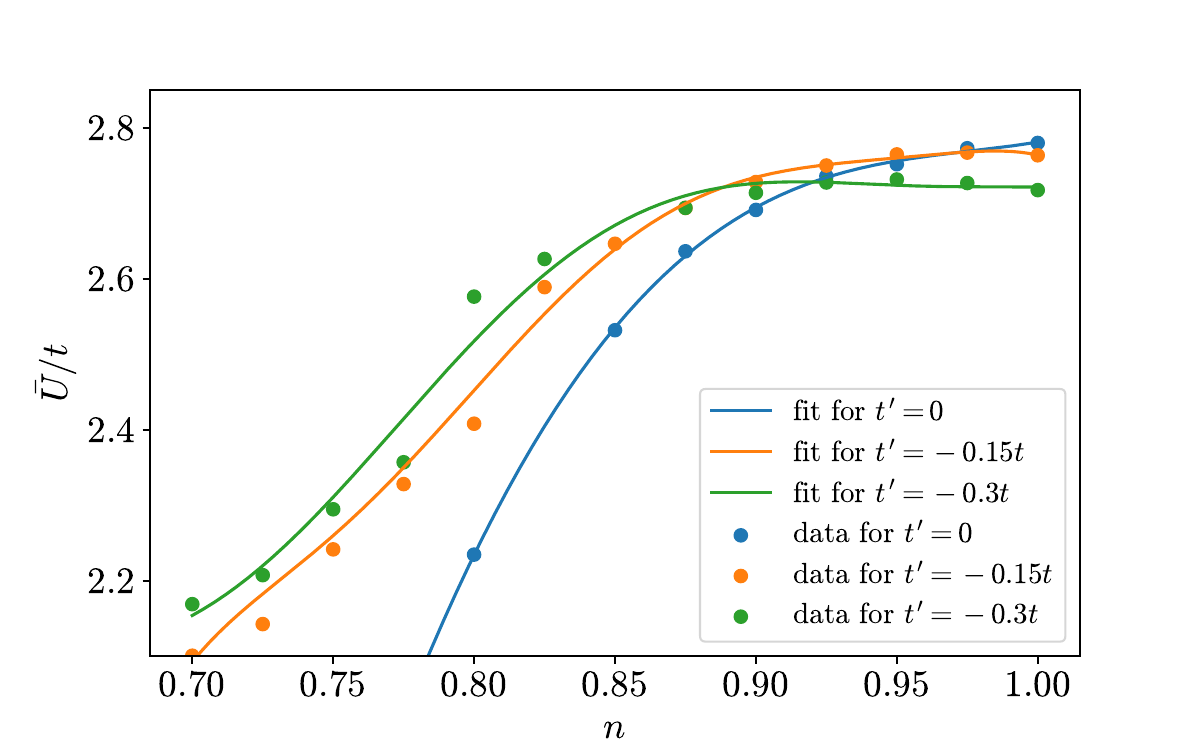}
\caption{Effective interaction $\bar U$ as a function of density for three choices of $t'$. The polynomial fit is used for $\bar U(n)$ in all subsequent results. The bare interaction is $U=5t$ in all plots.}
\label{fig: U(n)}
\end{figure}

We now present and discuss results obtained from the SU(2) gauge theory of fluctuating stripe order for some specific choices of model parameters. Defining the nearest neighbor hopping $t$ as our energy unit, we consider three choices for the next-to-nearest neighbor hopping: $t' = 0$,  $t' = -0.15t$ and $t' = -0.3t$. The first one, corresponding to pure nearest neighbor hopping, is not adequate for cuprate superconductors, but has nevertheless been a popular choice for numerical simulations of the Hubbard model, favoring especially stripe order \cite{Zheng2017, Huang2018, Qin2020, Xu2022}.
A next-nearest neighbor hopping $t' = -0.15t$ is frequently used to model the band structure of the LSCO family of cuprate superconductors, and $t' = -0.3t$ for YBCO \cite{Pavarini2001}.
For the Hubbard interaction we choose a moderate value $U = 5t$.
Applying the functional RG flow as described in Sec.~\ref{sec: chargons}, the corresponding effective interactions $\bar U(n)$ are about one half of the bare values, see Fig.~\ref{fig: U(n)}.
We choose the nearest neighbor hopping amplitude $t$ as our energy unit, setting $t=1$ in the figures.

Our main result is the spectral function for single-particle excitations, as obtained from the imaginary part of the electron Green function. To understand this result, we first discuss its ingredients, namely the stripe order of the chargons and the corresponding pseudospin stiffnesses.


\subsection{Stripe ordered chargons}

The chargons are treated in renormalized mean-field theory, see Sec.~\ref{sec: chargons}. We ignore superconductivity and focus on regions in the phase diagram where unidirectional stripe order minimizes the free energy. Pseudogap behavior in regions with N\'eel or spiral order has already been analyzed previously within our SU(2) gauge theory of fluctuating magnetic order \cite{Bonetti2022a, Forni2025}.
For $t'=0$, the mean-field ground state is N\'eel ordered only at half-filling, and stripe ordered away from half-filling, while for $T>0$ N\'eel and also spiral order are stabilized in a region around half-filling \cite{Scholle2023,Scholle2024}.
For sizable values of $t'/t$, the magnetic order is of stripe-type only for a sizable hole doping below half-filling ($n<1$), while for electron doping ($n>1$) and small hole doping, the magnetic order is of N\'eel-type or spiral, respectively \cite{Scholle2023,Scholle2024}.

As explained in Sec.~\ref{sec: stripe order}, we need to assume a {\em commensurate}\/ stripe order with a finite period $p$ in our numerical calculations. The order is then characterized by real space spin and charge profiles, which are described by finite Fourier series with wave vectors $\bQ_n = (2\pi n/p,n\pi)$ with $n \in \{0,1,...,P-1\}$, where $P=p$ if $p$ is even, and $P=2p$ if $p$ is odd. The spin profile is described by Fourier coefficients $M_n$ with odd $n$, and the charge profile by coefficients $\rho_n$ with even $n$.
The gap equations determine the Fourier coefficients $M_n$ and $\rho_n$ for any given period $p$, and we select the optimal period $p$ by minimizing the free energy with respect to $p$,
offering the following choices: $p \in \{ 2,4,\dots,32 \} \cup \{ 3,5,\dots,15 \}$, such that $P \leq 32$. Momentum integrals are replaced by discrete momentum sums corresponding to very large lattices (several hundred sites in each direction) with periodic boundary conditions and linear dimensions in $x$ direction which are integer multiples of $p$.

\begin{figure}[tb]
\centering
\includegraphics[width=16cm]{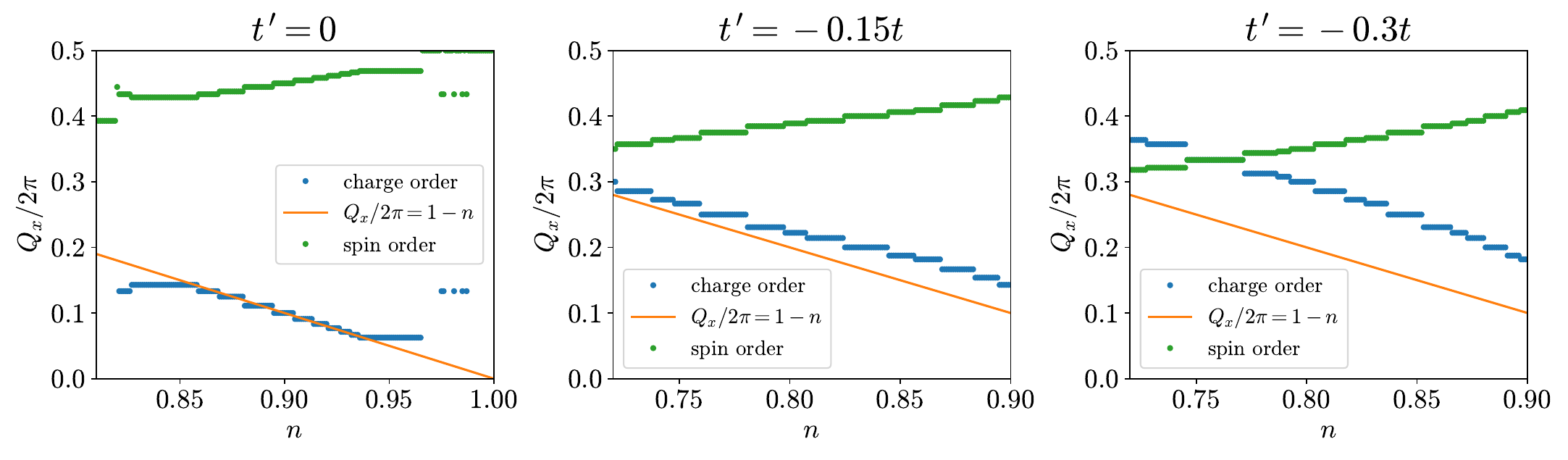}
\caption{$Q_x$ components of the dominant charge and spin ordering wave vectors as a function of the electron density $n$ for $U=5t$ and $T=0.02t$. Left: $t'=0$, center: $t'=-0.15t$, right: $t'=-0.3t$.}
\label{fig: dominant Q}
\end{figure}
The spin and charge profiles are never strictly harmonic, but rather a superposition with several wave vectors $\bQ_n$. We refer to the wave vectors with the largest Fourier coefficients $|M_n|$ and $|\rho_n|$ as the {\em dominant}\/ spin and charge order wave vectors $\bQ^s$ and $\bQ^c$, respectively. Strictly speaking there are inversion-symmetry related pairs of dominant wave vectors with equally maximal $|M_n|$ and $|\rho_n|$ for spin and charge, respectively. The $y$ component of $\bQ^s$ and $\bQ^c$ is always $\pi$ and $0$, respectively. The choice of $\bQ^s$ and $\bQ^c$ becomes unique if we consider only wave vectors with an $x$ component between $0$ and $\pi$.

In Fig.~\ref{fig: dominant Q} we show the $x$ components of the dominant ordering wave vectors $\bQ^s$ and $\bQ^c$ with $0 \leq Q_x^{s,c} < \pi$ as a function of the electron density at a fixed low temperature $T=0.02t$ for three choices of $t'$.
The dominant charge order wave vector exhibits an almost linear density dependence in a broad density range. Its form of a discontinuous step function is an artifact of the limited number of periodicities $p$ we can offer. For $t'=0$ the dominant charge order wave vector closely follows the line $Q_x = 2\pi(1-n)$, while for $t'=-0.15$ and $t'=-0.3t$ there is a constant offset. The $x$ component of $\bQ^s$ also varies roughly linearly with density. Deviations from the linear behavior and irregularities near the boundaries of the stripe region, especially for $t'=0$ near half-filling, are due to our restrictions of $p$.

\begin{figure}[tb]
\centering
\includegraphics[width=5.4cm]{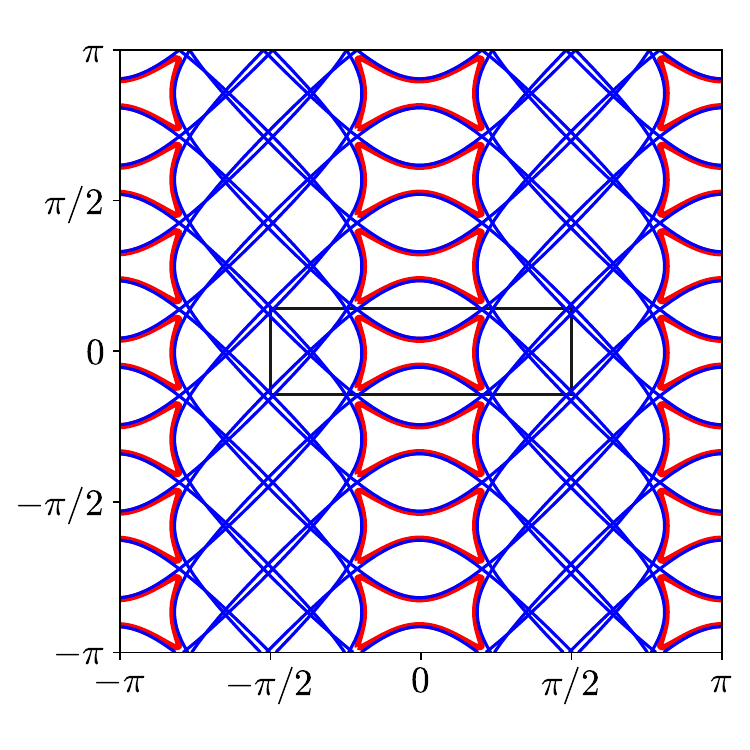}
\includegraphics[width=5.4cm]{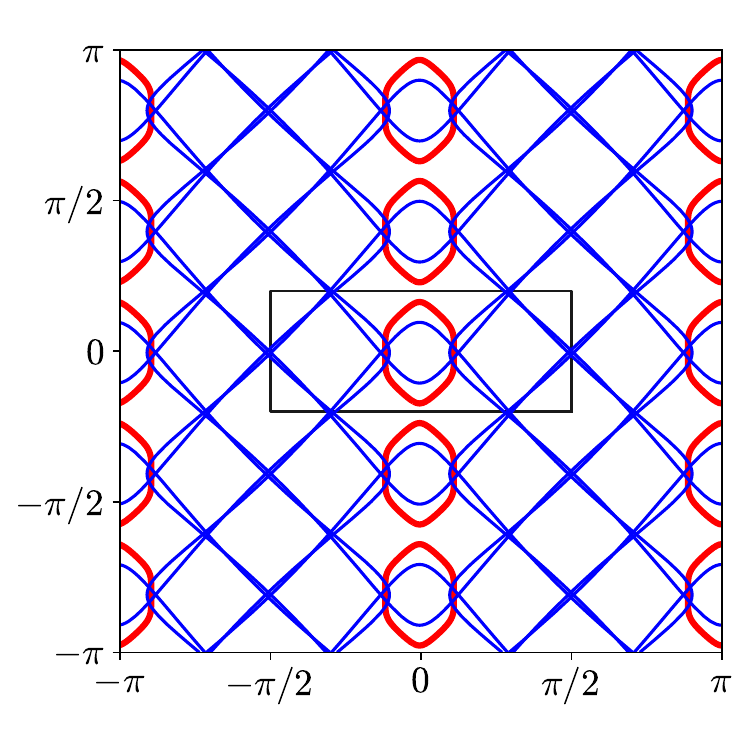}
\includegraphics[width=5.4cm]{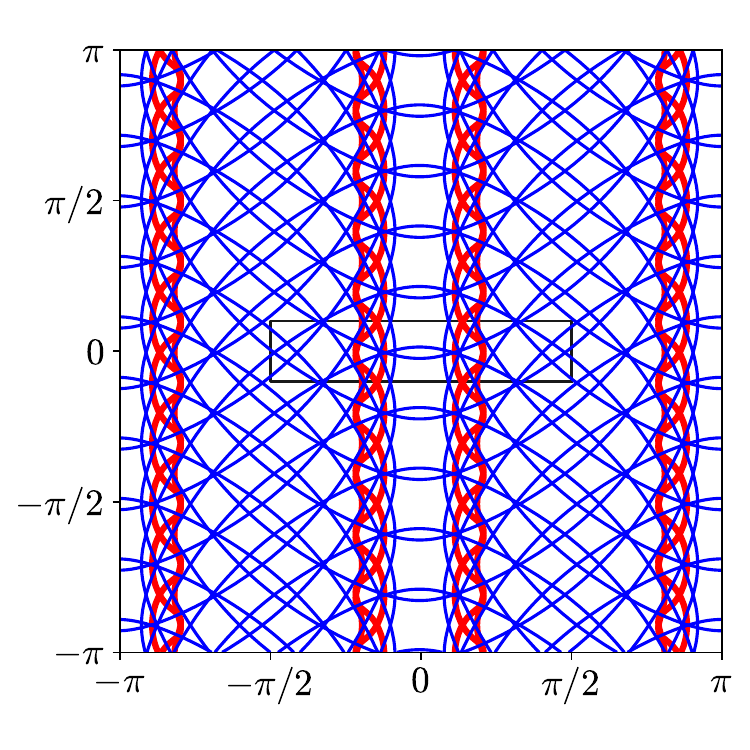}
\caption{Quasi-particle Fermi surfaces (red lines) of three stripe states with periodicities $p=7$ (left), $p=5$ (center), and $p=20$ (right), presented in the full Brillouin zone (repeated zone scheme). The Fermi surface of the corresponding non-interacting system (blue lines) is shown in the same repeated zone scheme for comparison. The black box in the center indicates the (reduced) magnetic Brillouin zones corresponding to the reduced translation invariance of the stripe ordered state.
Parameters: $T=0.02t$ and $t'=0$, $n=0.83$ (left), $t'=-0.15t$, $n=0.83$ (center), $t'=-0.3t$, $n=0.801$ (right).}
\label{fig: Fermi surfaces}
\end{figure}
In Fig.~\ref{fig: Fermi surfaces} we show the quasi-particle Fermi surfaces defined as zeros of the quasi-particle bands $E_\bk^\ell$ in the full Brillouin zone (repeated zone scheme) for three choices of $t'$ with three distinct stripe periodicities $p$. The Fermi surface of the corresponding non-interacting system is shown in the same repeated zone scheme for comparison. It consists of the original bare Fermi surface and its copies obtained from a shift by $\bQ_n$ with $n = 1,\dots,P-1$. Some of the quasi-particle bands $E_\bk^\ell$ are completely filled, and others are completely empty. Hence, generally only some of the bands are {\em partially}\/ filled and form a Fermi surface sheet. A large magnetic gap favors a reduced number of partially filled bands (while most bands are either completely filled or empty). In the left and center panels of Fig.~\ref{fig: Fermi surfaces}, only one of the quasi-particle bands is partially filled, leading to a relatively simple Fermi surface topology with a single hole (left) and electron (center) pocket in the reduced Brillouin zone.
In the right panel, three of the twenty quasi-particle bands are partially filled, and the Fermi surface consists of various open lines, as if the system was quasi one-dimensional. For a slightly higher filling (not shown), pockets appear in addition to the open lines.
Quite generally, the shape of the Fermi surfaces can change significantly upon small changes of the density.


\subsection{Stiffnesses}

\begin{figure}[tb]
\centering
\includegraphics[width=8cm]{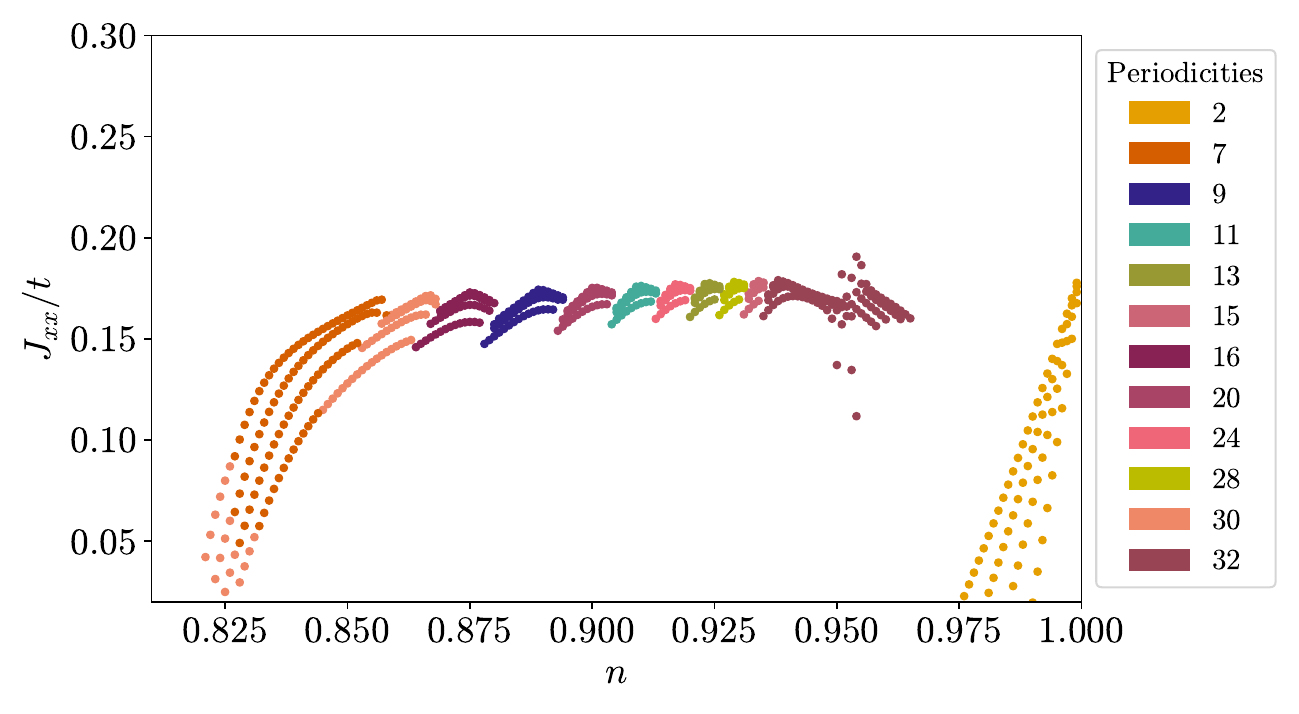}
\includegraphics[width=8cm]{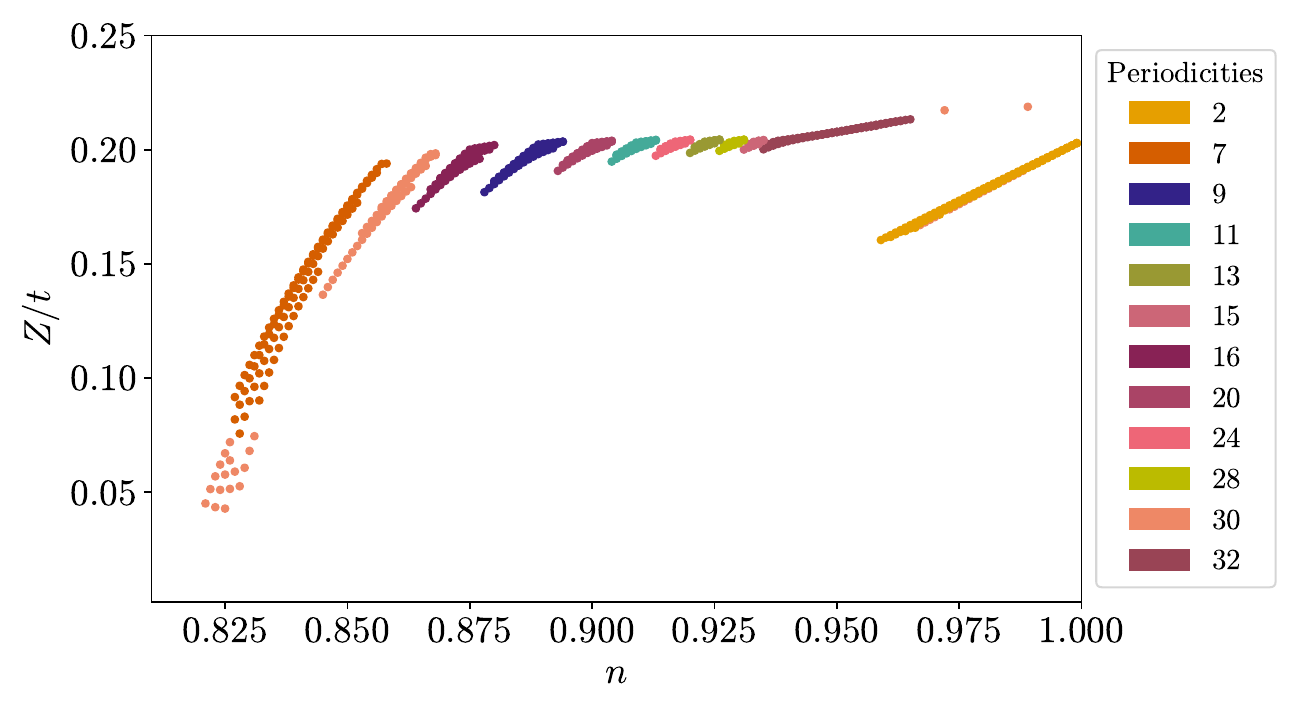}
\includegraphics[width=8cm]{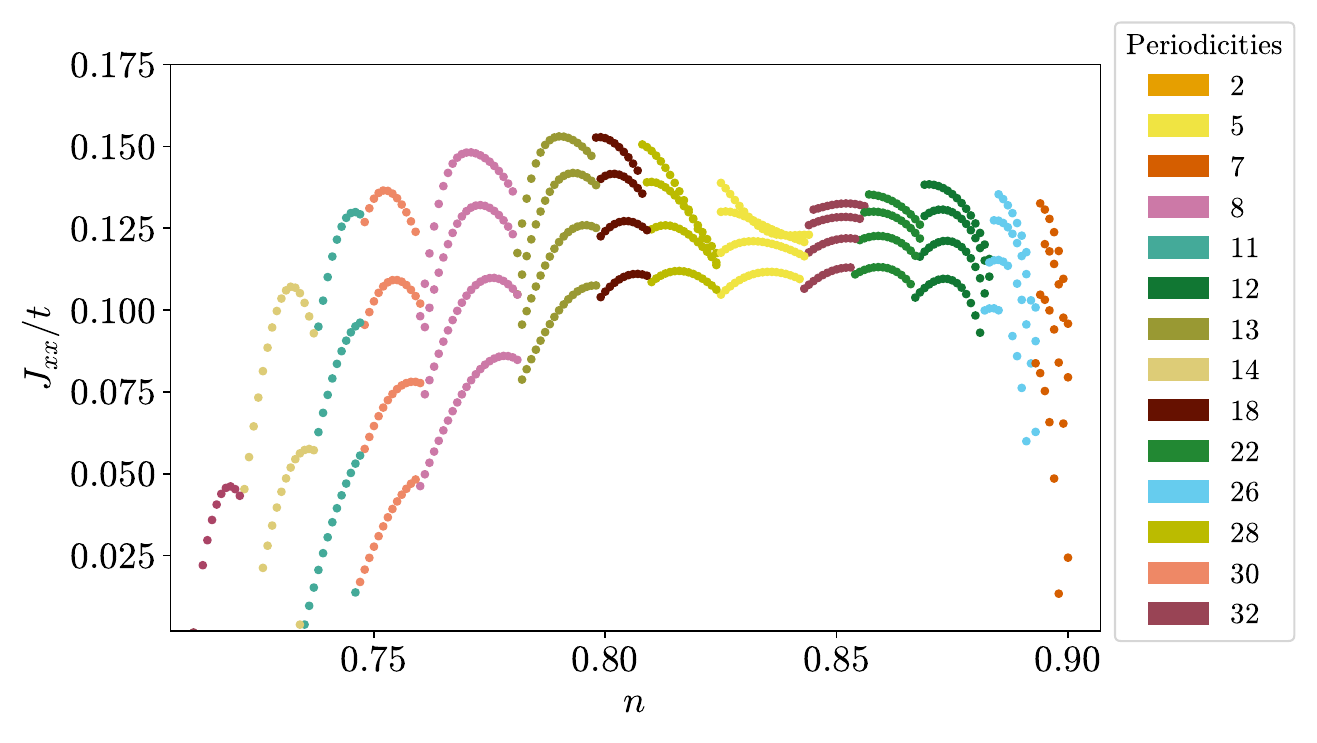}
\includegraphics[width=8cm]{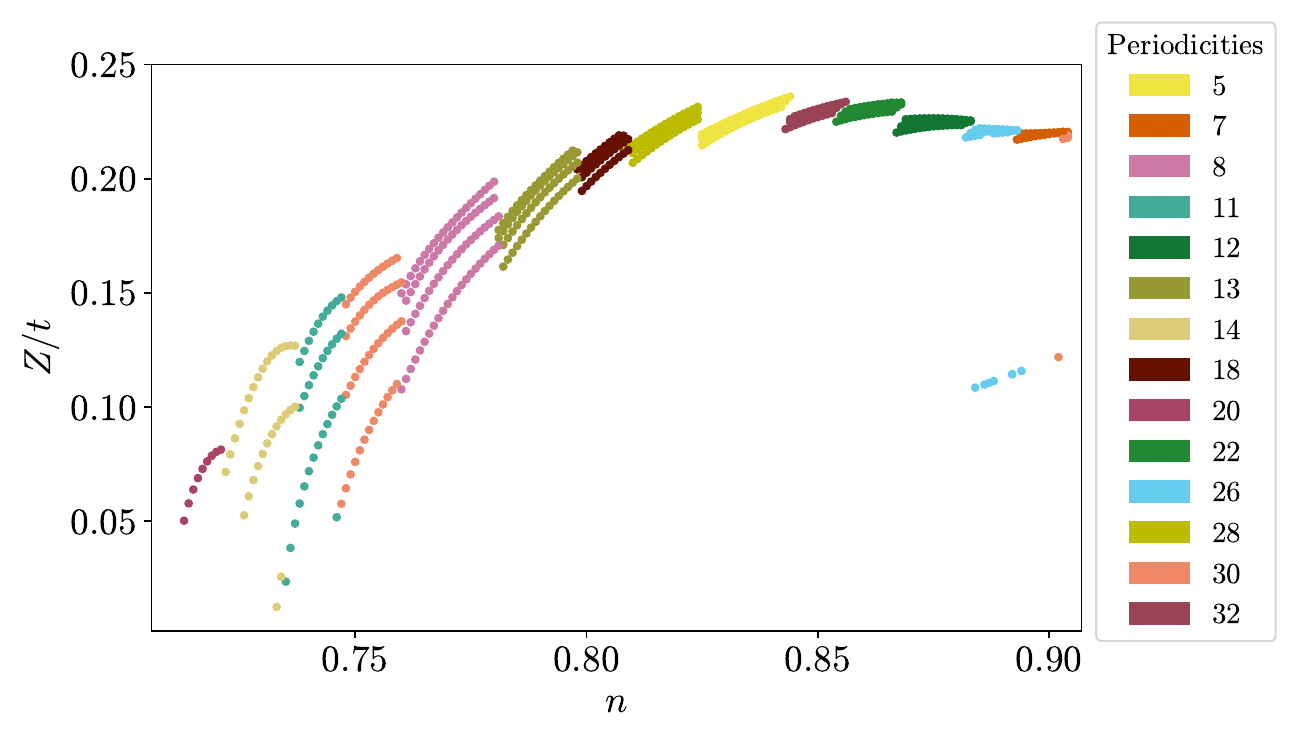}
\includegraphics[width=8cm]{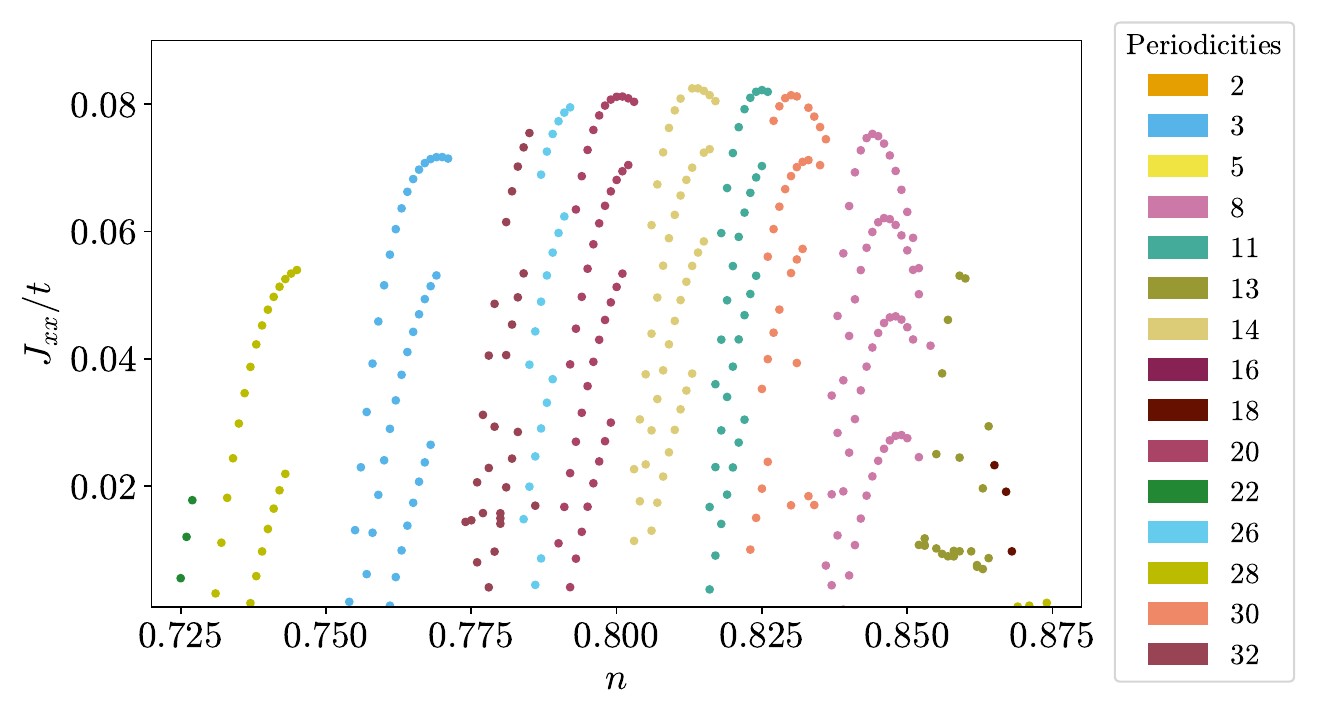}
\includegraphics[width=8cm]{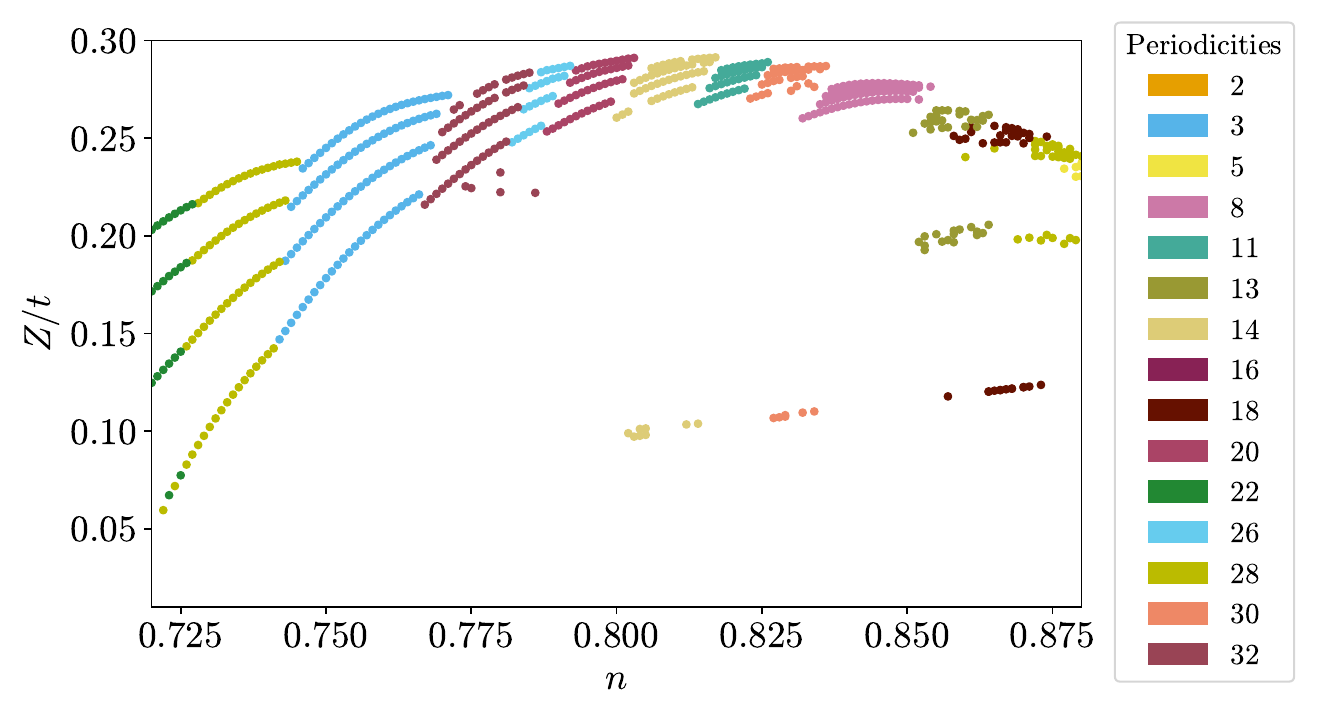}
\caption{Spatial stiffness in $x$ direction $J_{xx}$ and temporal stiffness $Z$ as functions of density in the stripe ordered regime for $t'=0$ (top), $t'=-0.15t$ (center), and $t'=-0.3t$ (bottom) at various low temperatures $T = 0.02t, 0.03t, 0.04t, 0.05t$. The periodicities $p$ of the stripe order which minimize the free energy are indicated by different colors. Larger stiffnesses (at a given density) correspond to lower temperatures.}
\label{fig: stiffness}
\end{figure}

In Figs.~\ref{fig: stiffness} we show the spatial stiffness in $x$ direction $J_{xx}$ and the temporal stiffness $Z$ as functions of density in the stripe ordered regime for various low temperatures, $U=5t$, and three choices of $t'$.
The spatial stiffness in $y$ direction $J_{yy}$ behaves similarly. The indicated periodicities $p$ minimize the free energy within the set
$p \in \{ 2,4,\dots,32 \} \cup \{ 3,5,\dots,15 \}$.
We show the stiffnesses only in the density interval in which stripe order is the energetically most favorable magnetic order.
The spatial stiffness for fixed $p$ drops rapidly away from its maximum as a function of density.
For $t'=0$ and $t'=-0.3t$, there are densities for which none of the available periodicities lead to a state with a positive spatial stiffness.
This is due to the limited set of periodicities we can offer, which limits the set of allowed wave vectors $\bQ_n$.
Discarding the irregularities resulting from this restriction, one can see that the highest data points (for each temperature) in the figures follow regular envelopes, which define the stiffnesses one should obtain in the thermodynamic limit, if arbitrary periodicities were allowed. The relatively low spatial stiffness near $n \approx 0.83$ for $t'=-0.15t$ may be associated with the van Hove singularity in this regime.

The size of the stiffnesses (deep inside the stripe regime) is of the same order of magnitude as the ones in the N\'eel state at half-filling \cite{Bonetti2022, Vilardi2025}.
The stiffnesses consist of a ``bare'' contribution and an interaction correction (see Secs.~\ref{sec: bare contributions} and \ref{sec: interaction correction}). The latter, which is much harder to compute, is generally much smaller than the former.


\subsection{Electron spectral function}

We now present results for the spectral function for single-particle excitations, which is directly related to the imaginary part of the electron Green function by
\begin{equation}
 A(\bk,\bk',\omega) =
 - \frac{1}{\pi} {\rm Im} G^e(\bk,\bk',\omega+i0^+) \, .
\end{equation}
We focus on its momentum diagonal part, which, via Eq.~\eqref{eq: Ge final}, can be written as
\begin{eqnarray} \label{eq: spec fct}
 A(\bk,\omega) &=& - \frac{1}{\pi} {\rm Im} G^e({\bf 0},{\bf 0};\bk,\omega) \nonumber \\
 &=& \int_\bq \sum_{\ell} \sum_{p=\pm} \frac{1}{2Z \omega_\bq} \,
 \left[ b(\omega_\bq) + f(pE_{\bk-\bq}^\ell) \right] (g_{\bk-\bq}^\ell)_{00} \,
 \delta(\omega + i0^+ - E_{\bk-\bq}^\ell + p\,\omega_\bq) \, . \hskip 5mm
\end{eqnarray}
$A(\bk,\omega)$ can be measured by angular resolved photoemission (ARPES) \cite{Damascelli2003}.

\begin{figure}[tb]
\centering
\includegraphics[width=5.4cm]{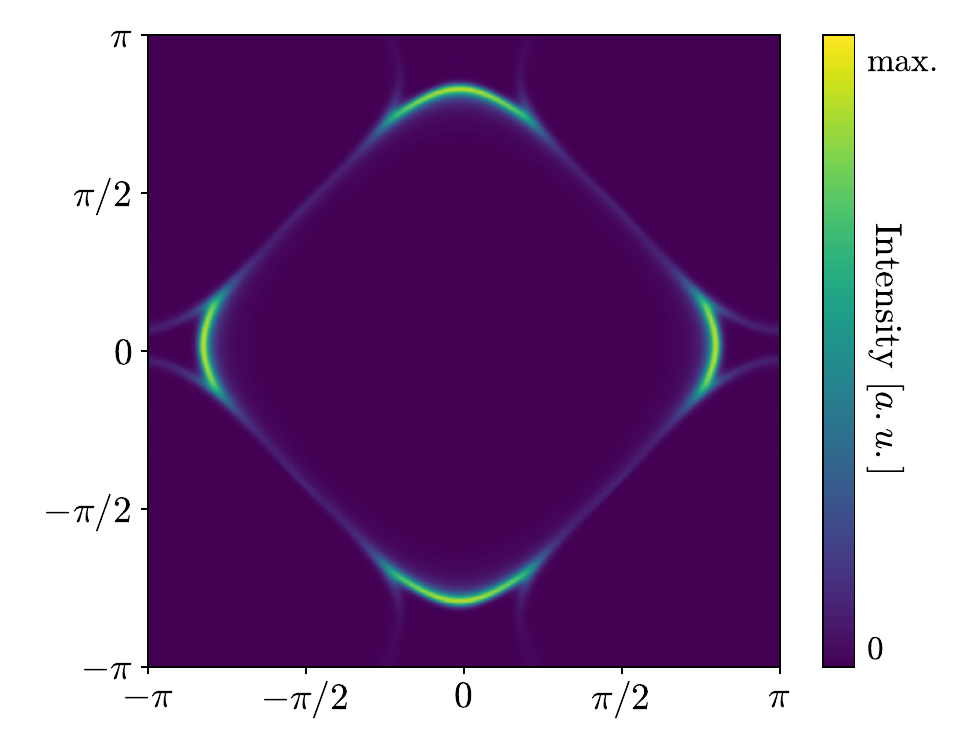}
\includegraphics[width=5.4cm]{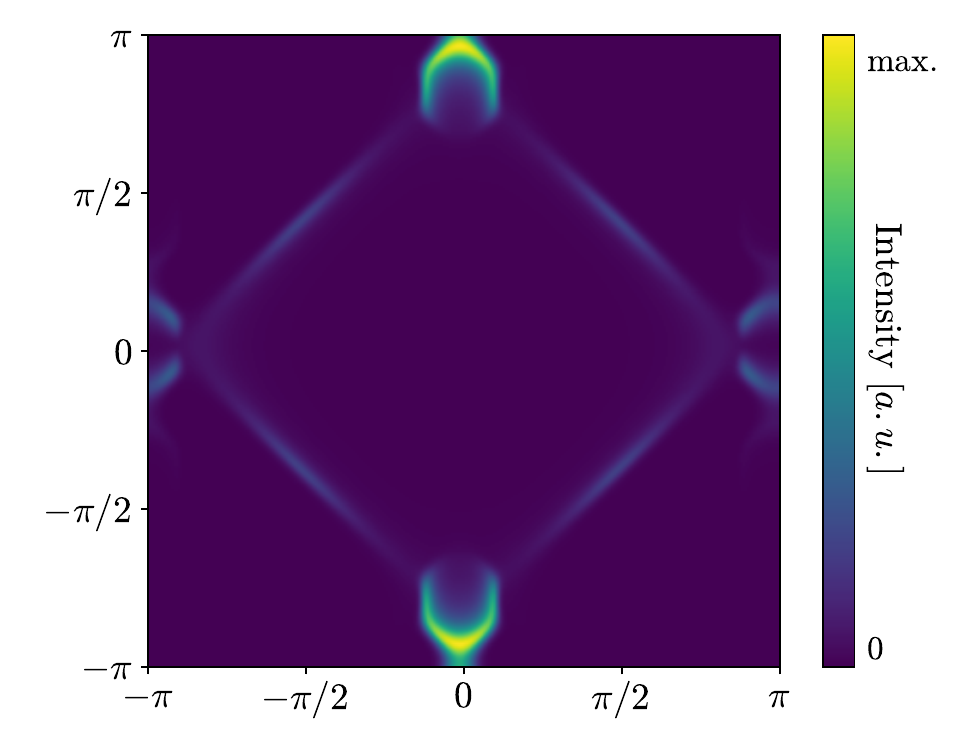}
\includegraphics[width=5.4cm]{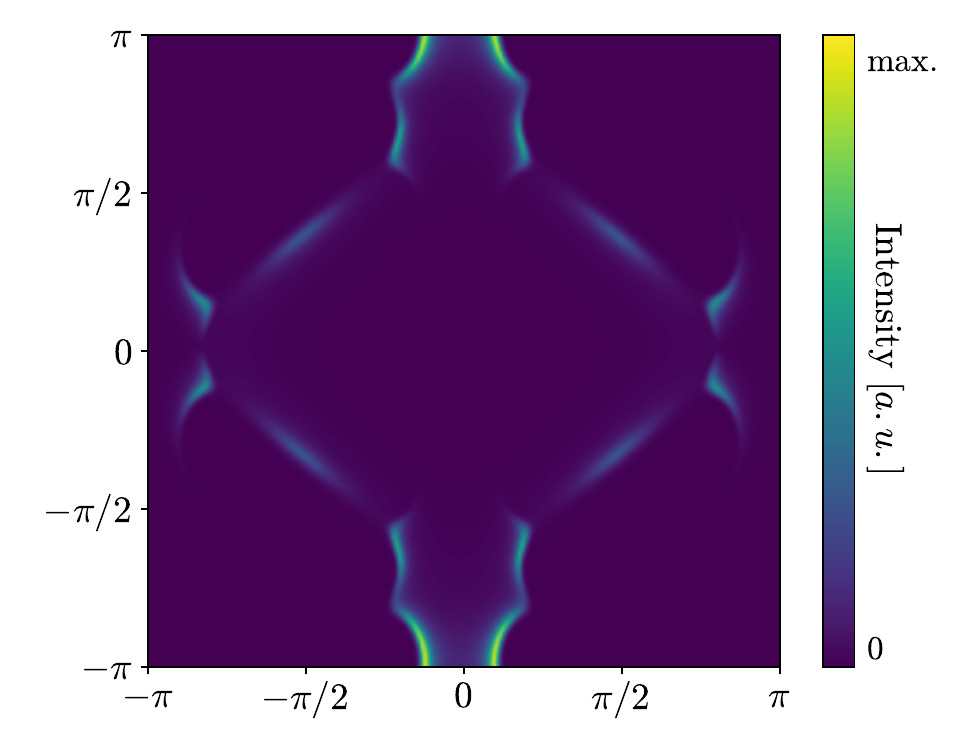}
\caption{Spectral function for single-particle excitations $A(\bk,\omega)$ at $\omega=0$. Parameters: $T=0.02t$ and $t'=0$, $n=0.83$ (left), $t'=-0.15t$, $n=0.83$ (center), $t'=-0.3t$, $n=0.801$ (right).}
\label{fig: spec fct}
\end{figure}

In Fig.~\ref{fig: spec fct} we show the spectral function $A(\bk,\omega)$ at $\omega=0$ for three distinct cases of fluctuating stripe order. The parameters are the same as in Fig.~\ref{fig: Fermi surfaces}.
In a normal Fermi liquid, $A(\bk,0)$ exhibits a sharp peak at the Fermi surface. In a state with magnetic and/or charge order, $A(\bk,0)$ is peaked at the (reconstructed) Fermi surface of the quasi-particles. The size of the peaks is strongly momentum dependent, due to weight factors which are very small for momenta $\bk$ remote from the bare Fermi surface \cite{Eberlein2016}. The spinon fluctuations broaden the peaks to some extent.
Indeed, the spectral functions $A(\bk,0)$ in Fig.~\ref{fig: spec fct} exhibit peaks on those segments of the quasi-particle Fermi surfaces which are close to the bare Fermi surface.
Some of these segments have the form of Fermi arcs, which, depending on the parameters, can be situated near the nodal directions (near the Brillouin zone diagonal) or in the antinodal region.

Our result for the spectral function for $t'=-0.15t$, with a large spectral weight in the antinodal regions, and more faint straight lines connecting these regions, resembles observations from ARPES in ``stripy'' cuprates, namely $\rm La_{1.4}Nd_{0.6}Sr_xCuO_4$ and $\rm La_{1.85}Sr_{0.15}CuO_4$ \cite{Zhou2001}. Hence, these experimental observations can be naturally explained by fluctuating stripe order.
Nowhere in the stripe ordered regime we find an exclusive collection of four nodal Fermi arcs as observed experimentally, in particular, in the pseudogap regime of cuprates with Bismuth, such as $\rm Bi_2Sr_2CaCu_2O_{8+\delta}$ \cite{Damascelli2003}. Such nodal Fermi arcs are obtained theoretically in a state with fluctuating N\'eel or spiral magnetic order \cite{Bonetti2022a}. Hence, pseudogap behavior with four nodal Fermi arcs must be due to fluctuating N\'eel or spiral order rather than fluctuating stripe order.


\section{Conclusions}

We have presented an SU(2) gauge theory of fluctuating stripe order in the two-dimensional Hubbard model. Our theory is based on a fractionalization of the electron operators in a fermionic chargon with a pseudospin degree of freedom, and a charge-neutral spinon capturing fluctuations of the spin orientation \cite{Schulz1995, Dupuis2002, Borejsza2004, Sachdev2009}. The chargons are treated in a renormalized mean-field theory with an effective interaction obtained from a renormalization group flow. We have focused on regions in the phase diagram where the chargons order in a spin-charge stripe pattern, which occurs generically at sufficiently large hole-doping \cite{Scholle2023, Scholle2024}. Fluctuations of the spin orientation are described by a nonlinear sigma model. The parameters of the sigma model, the spatial and temporal pseudospin stiffnesses, have been computed from a renormalized RPA. Our approximations are applicable for a moderate Hubbard interaction.
The spinon fluctuations prevent breaking of the physical SU(2) spin symmetry of the Hubbard model at any finite temperature, in agreement with the Mermin-Wagner theorem, resulting in a charge ordered pseudogap phase with a reconstructed Fermi surface and a spin gap.

Concrete solutions have been obtained for a moderate Hubbard interactions $U=5t$.
Solving the renormalized mean-field equations we have computed the spin and charge amplitudes defining the stripe profile. The bare dispersion is split into numerous quasi-particle subbands, their number being proportional to the periodicity of the stripe pattern. Correspondingly, the Fermi surface is fragmented into a collection of electron and/or hole pockets and open lines. While the stripe profile (in real space) is never strictly harmonic, it is usually dominated by one Fourier component, that is, a single wave vector. The charge order wave vector and the deviation of the spin order wave vector from $(\pi,\pi)$ increase monotonically and almost linearly with hole-doping, and only along one of the momentum axes ($q_x$ or $q_y$).

Small electron pockets have been identified in regions of (fluctuating) charge order of cuprates via quantum oscillation and transport experiments \cite{Proust2019}. To obtain such pockets from charge order alone requires a superposition of multiple (non-collinear) ordering wave vectors \cite{Harrison2011}. Stripe order offers another natural mechanism for electron pockets, where a single (uni-unidirectional) wave vector is sufficient.

Deep inside the stripe regime in the phase diagram, the spatial and temporal pseudospin stiffnesses are of the same order of magnitude as for the N\'eel state at half-filling. Both quantities decrease upon approaching the edge of the magnetically ordered region at large hole-doping, and also upon increasing the temperature. In the ground state, the stiffnesses determine the stability of the magnetically ordered mean-field state with respect to quantum fluctuations. Since the ground state of the Hubbard model at half-filling is known to exhibit antiferromagnetic long-range order from (sign-free) Monte Carlo calculations, the sizable stiffnesses in the stripe ordered hole-doped regime indicate that the magnetic order may be robust with respect to fluctuations at $T=0$. Indeed, a stripe ordered ground state has been found for several sizable values of the hole-doping in exact numerical simulations \cite{Zheng2017, Huang2018, Qin2020}.

We have computed the spectral function $A(\bk,\omega)$ for single-particle excitations, a quantity that can be measured by ARPES \cite{Damascelli2003}. In a normal Fermi liquid, $A(\bk,0)$ exhibits a pronounced peak at the Fermi surface. In a state with magnetic and/or charge order, $A(\bk,0)$ is peaked at the (reconstructed) Fermi surface of the quasi-particles. The size of the peaks is strongly momentum dependent due to weight factors which are very small for momenta $\bk$ remote from the bare Fermi surface \cite{Eberlein2016}. The spinon fluctuations broaden the peaks to some extent. Indeed, the spectral function $A(\bk,0)$ obtained in our theory of fluctuating stripe order exhibits pronounced peaks on those parts of the quasi-particle Fermi surfaces which are close to the bare Fermi surface.
Some of these parts have the form of Fermi arcs, but not necessarily centered around the nodal directions (near the Brillouin zone diagonal).
For $t'=-0.15t$, applicable to the LSCO family of cuprates, we obtain a large spectral weight in the antinodal regions, and more faint straight lines connecting these regions. This resembles observations from ARPES in $\rm La_{1.4}Nd_{0.6}Sr_xCuO_4$ and $\rm La_{1.85}Sr_{0.15}CuO_4$ \cite{Zhou2001}, where static or fluctuating stripes are known to play an important role.
Nowhere in the stripe ordered regime we have found a simple collection of four nodal Fermi arcs as observed experimentally in the pseudogap regime of other cuprates, especially from the Bismuth family \cite{Damascelli2003}, and found theoretically in a state with fluctuating N\'eel or spiral magnetic order \cite{Bonetti2022a}. Hence, we conclude that pseudogap behavior with four nodal Fermi arcs must be due to fluctuating N\'eel or spiral order rather than fluctuating stripe order.
Note, however, that finite-energy dynamical stripe or pure charge order fluctuations do not necessarily affect the Fermi surface obtained from the spectral function at zero energy \cite{Grilli2009}. Hence, the observation of fluctuating stripes from other experimental probes in cuprates \cite{Kivelson2003} may still be reconciled with purely nodal Fermi arcs seen in ARPES.

While the spinon fluctuations preserve the physical SU(2) spin symmetry, at least at finite temperatures, the charge order and nematicy associated with the stripe order of the chargons remains. Commensurate charge order is not excluded by the Mermin-Wagner theorem, as it breaks only a discrete, not a continuous symmetry. Charge order has indeed been observed in some cuprate compounds in the presence of a strong magnetic field, which suppresses the competing superconductivity \cite{Wu2011, Chang2012}, and the ordering wave vector is oriented along the $q_x$ or $q_y$ axes, in agreement with our results.

The treatment of the chargons by renormalized mean-field theory is applicable only for a weak or moderate Hubbard interaction. Our theory can be extended to strong coupling by computing the chargon order parameter and the corresponding pseudospin stiffnesses from dynamical mean-field theory.

\newpage


\section*{Acknowledgements}

We thank Demetrio Vilardi for computing the renormalized Hubbard interaction, and for valuable comments on our manuscript.
We are grateful to Andrey Chubukov, Jakob Dolgner, Antoine Georges, Marco Grilli, Subir Sachdev, Mathias Scheurer, Robin Scholle, Oleg Sushkov, and Hiroyuki Yamase for valuable discussions.
P.M.B.\ acknowledges support by the German National Academy of Sciences Leopoldina through Grant No.\ LPDS 2023-06 and the Gordon and Betty Moore Foundation’s EPiQS Initiative Grant GBMF8683.


\appendix


\section{Alternative Ward identity} \label{app: alternative WI}

Another Ward identity holds for a {\em modified}\/ gauge kernel $\bar K_{\mu\nu}^{ab}$, which is related to $K_{\mu\nu}^{ab}$ by a Legendre transform. In Ref.~\cite{Bonetti2022ward} it was assumed that $\bar K_{\mu\nu}^{ab} = K_{\mu\nu}^{ab}$, but later it was shown that they actually differ \cite{Bonetti2024,Goremykin2024}.
In a magnetic state with an arbitrary spin order $\bra \bS(x) \ket = {\bf m}(\br)$, the modified gauge kernel obeys the Ward identity \cite{Bonetti2022ward,Goremykin2024}
\begin{equation} \label{eq: wardid2a}
 \partial_{\mu} \partial'_\nu \bar K_{\mu\nu}^{aa'}(x,x') =
 \eps^{abc} \eps^{a'b'c'} C^{bb'}(x,x') m^c(\br) m^{c'}(\br') \, ,
\end{equation}
where $C^{bb'}(x,x')$ is the {\em inverse}\/ spin susceptibility, related to $\chi^{aa'}(x,x')$ by
\begin{equation}
 \int_{x''} C^{aa''}(x,x'') \chi^{a''a'}(x'',x') = \delta_{aa'} \delta(x-x') \, .
\end{equation}
Unlike the Ward identity presented in the main text, this identity does not involve an external symmetry breaking field.
For collinear stripe order with a spin profile $S^1(\br)$ in $x$-direction, the Ward identity reads
\begin{equation} \label{eq: wardid2b}
 \partial_\mu \partial'_\nu \bar K_{\mu\nu}^{aa'}(x,x') =
 \left\{ \begin{array}{lll}
 {\phantom-} C^{\bar a \bar a}(x,x') m^1(\br) m^1(\br') &
 \mbox{for} & a = a' \in \{2,3\} \\
 - C^{\bar a \bar a'}(x,x') m^1(\br) m^1(\br') &
 \mbox{for} & a \neq a' \in \{2,3\} \\
 0 & \mbox{else} &
 \end{array} \right. \, ,
\end{equation}
with $\bar 2 = 3$ and $\bar 3 = 2$.
Defining a {\em weighted}\/ inverse spin correlation function as
\begin{equation} \label{eq: def CSS}
 \Ct^{aa'}(x,x') =
 \frac{1}{m^2} m^1(\br) m^1(\br') \, C^{aa'}(x,x') \, ,
\end{equation}
and Fourier transforming, yields
\begin{equation} \label{eq: wardid2c}
 (\bq,iq_0)_\mu (\bq',iq'_0)_\nu \bar K_{\mu\nu}^{aa'}(q,q') =
 \left\{ \begin{array}{lll}
 {\phantom-} m^2 \Ct^{\bar a \bar a}(q,q') & \mbox{for} & a = a' \in \{2,3\} \\
 - m^2 \Ct^{\bar a \bar a'}(q,q') & \mbox{for} & a \neq a' \in \{2,3\} \\
 0 & \mbox{else} &
 \end{array} \right. \, .
\end{equation}

We define modified stiffnesses in analogy to Eq.~\eqref{eq: spatial stiffness} and \eqref{eq: temporal stiffness} as
\begin{eqnarray}
 \bar J_{\alf\beta}^{ab} &=&
 \lim_{\bq \to 0} \bar K_{\alf\beta}^{ab}(\bq,0) \, , \\
 \bar Z^{ab} &=&
 - \lim_{q_0 \to 0} \bar K_{00}^{ab}({\bf 0},0) \, ,
\end{eqnarray}
where $\bar K_{\alf\beta}^{ab}(q) = \bar K_{\alf\beta}^{ab}(q,q)$.
The Ward identity Eq.~\eqref{eq: wardid2c} then yields the relations
\begin{eqnarray}
 \label{eq: wardstiff2a}
 \bar J_{\alf\beta}^{aa} &=&
 \frac{m^2}{2} \left. \partial_{q_\alf} \partial_{q_\beta}
 \Ct^{\bar a \bar a}(q,q) \right|_{q=0} \, , \\
 \label{eq: wardstiff2b}
 \bar Z^{aa} &=&
 \frac{m^2}{2} \left. \partial_{q_0}^2
 \Ct^{\bar a \bar a}(q,q) \right|_{q=0} \, .
\end{eqnarray}

For a N\'eel or spiral state, the modified stiffnesses and the stiffnesses defined via $K_{\mu\nu}^{ab}$ in an infinitesimal external field are actually identical, that is, $\bar J_{\alf\beta}^{aa} = J_{\alf\beta}^{aa}$ and $\bar Z^{aa} = Z^{aa}$, as shown in Refs.~\cite{Bonetti2024,Goremykin2024}. We expect analogous relations to hold in the case of stripe order, too, but refrain from working out a detailed proof, because they are not relevant for our results.


\section{Cancellation of stiffness contributions} \label{app: cancellation}

Here we show that the diamagnetic and the bare paramagnetic contributions to the 11-component of the stiffness, $J_{\alf\beta}^{11}$, cancel each other. We start from the expression \eqref{eq: Kd2} for the diamagnetic contribution and rewrite it by a partial integration.
\begin{eqnarray}
 K_{\alf\beta}^{d,11} &=& \frac{T}{4} \sum_{k_0} \int'_\bk
 \tr \left[ \Gam_\bk^{d,\alf\beta} \cG(k) \right] =
 \frac{T}{2} \sum_{k_0} \int'_\bk
 \tr \left[ (\partial_{k_\beta} \Gam_\bk^{\alf,+} \cG_\sg(k) \right]
 \nonumber \\
 &=& - \frac{T}{2} \sum_{k_0} \int'_\bk
 \tr \left[ \Gam_\bk^{\alf,+} \partial_{k_\beta} (ik_0 - \cH_\bk)^{-1} \right]
 \nonumber \\
 &=& \frac{T}{2} \sum_{k_0} \int'_\bk
 \tr \left[ \Gam_\bk^{\alf,+} \, \cG_\sg(k) [\partial_{k_\beta} (ik_0 - \cH_\bk)]
 \cG_\sg(k) \right] \nonumber \\
 &=& - \frac{T}{2} \sum_{k_0} \int'_\bk
 \tr \left[ \Gam_\bk^{\alf,+} \, \cG_\sg(k) \Gam_\bk^{\beta,+} \, \cG_\sg(k) \right]
 = - J_{\alf\beta,0}^{p,11} \, .
\end{eqnarray}
Hence, $K_{\alf\beta}^{d,11} + J_{\alf\beta,0}^{p,11} = 0$.



\section{Symmetries of the bare susceptibility} \label{app: susc sym}

In the following we derive symmetry relations for the bare susceptibility
$\chi_0^{ab}(\bQ_n,\bQ_{n'};\bq,\omega)$, from which one can see which components vanish at
$\bq = \mathbf{0}$ or $\omega=0$.
Specializing Eq.~\eqref{eq: Kp0' b} to $\mu=\nu=0$, the bare susceptibility can be expressed as
\begin{equation} \label{eq: chi0 a}
 \chi_0^{ab}(\bQ_n,\bQ_{n'};\bq,\omega) =
 - \frac{1}{2P} \int_\bk \sum_{\ell,\ell'}
 A_{\ell\ell'}^{ab}(\bQ_n,\bQ_{n'};\bk,\bq) F_{\ell\ell'}(\bk,\bq,\omega) \, ,
\end{equation}
with $F_{\ell\ell'}(\bk,\bq,\omega)$ from Eq.~\eqref{eq: Fll'}, and
\begin{equation} \label{eq: cohfac}
 A_{\ell\ell'}^{ab}(\bQ_n,\bQ_{n'};\bk,\bq) =
 \frac{1}{2} \tr \left[ \Sg^a \Pi_n g_{\bk+\bq}^\ell \Pi_{n'}^T \Sg^b g_\bk^{\ell'}
 \right] \, ,
\end{equation}
where $\Pi_n g_{\bk+\bq}^\ell \Pi_{n'}^T$ and $g_\bk^{\ell'}$ are shorthand notations for the diagonal $2 \times 2$ block matrices in Eq.~\eqref{eq: Amunu}.
For the following steps it is convenient to extend the momentum integral over the full Brillouin zone. The prefactor $P^{-1}$ in Eq.~\eqref{eq: chi0 a} compensates the corresponding overcounting.

Let us consider the momentum space mapping $T: \bk \mapsto -\bk - \bQ_{P-1}$. In the case of N\'eel order this reduces to the mapping employed to derive symmetry relations in Ref.~\cite{Bonetti2022}. We define a corresponding matrix $\cT$ acting in the basis $\cB_\sg$,
\begin{equation}
 \cT = \left( \begin{matrix}
 & & 1 \\ & \iddots & \\ 1 & & \end{matrix} \right)
 \end{equation}
For inversion symmetric systems, $T$ maps the Hamiltonian matrix defined in
Eq.~\eqref{eq: HMF explicit} as $\cH_\bk \mapsto \cT \cH_{-\bk} \cT$, so that the matrices $g_\bk^\ell$ obey the relation
\begin{equation} \label{eq: g trans}
 g_{-\bk-\bQ_{P-1}}^\ell = \cT g_\bk^\ell \cT \, .
\end{equation}
The cyclic permutation matrix $\Pi_n$ transforms as $\cT \Pi_n \cT = \Pi_{P-n}$, and the ``extended'' Pauli matrices $\Sg^a$ as $\cT \Sg^a \cT = s^a \Sg^a$, with
\begin{equation}
 s^a = \left\{ \begin{array}{lll}
  1 & \mbox{for} & a = 0,1 \\
 -1 & \mbox{for} & a = 2,3
 \end{array} \right. \, .
\end{equation}
Moreover $(\Sg^a)^* = p^a \Sg^a$, where
\begin{equation}
 p^a = \left\{ \begin{array}{lll}
 -1 & \mbox{for} & a = 2 \\
  1 & \mbox{for} & a = 0,1,3
 \end{array} \right. \, .
\end{equation}

The matrices $g_\bk^\ell$ can be chosen real, so that
\begin{equation} \label{eq: A*1}
 \left[ A_{\ell\ell'}^{ab}(\bQ_n,\bQ_{n'};\bk,\bq) \right]^* =
 p^a p^b A_{\ell\ell'}^{ab}(\bQ_n,\bQ_{n'};\bk,\bq) \, .
\end{equation}
They are also symmetric, so that $(g_\bk^\ell)^T = g_\bk^\ell$, and
$(\Pi_n g_\bk^\ell \Pi_{n'}^T)^T = \Pi_{n'} g_\bk^\ell \Pi_n^T$.

Using the invariance of the trace under transposition or cyclic permutations of its argument, and $g_\bk^\ell$ being symmetric, one can show that
\begin{equation} \label{eq: A*2}
 \left[ A_{\ell\ell'}^{ab}(\bQ_n,\bQ_{n'};\bk,\bq) \right]^* =
 A_{\ell\ell'}^{ba}(\bQ_{n'},\bQ_n;\bk,\bq) \, .
\end{equation}
Combining Eqs.~\eqref{eq: A*1} and \eqref{eq: A*2}, we obtain
\begin{equation} \label{eq: A*3}
 A_{\ell\ell'}^{ab}(\bQ_n,\bQ_{n'};\bk,\bq) =
 p^a p^b A_{\ell\ell'}^{ba}(\bQ_{n'},\bQ_n;\bk,\bq) \, .
\end{equation}

Using Eq.~\eqref{eq: g trans} and $\Pi_n \Sg^a \Pi_n^T = (s^a)^n \Sg^a$, one can derive the identity
\begin{equation}
 A_{\ell'\ell}^{ab}(\bQ_n,\bQ_{n'};-\bk-\bQ_{P-1}-\bq,\bq) =
 (s^a)^n (s^b)^{n'} s^a s^b A_{\ell\ell'}^{ba}(\bQ_{n'},\bQ_n;\bk,\bq) \, .
\end{equation}
Combining this with Eq.~\eqref{eq: A*3}, we obtain
\begin{equation} \label{eq: A sym}
A_{\ell'\ell}^{ab}(\bQ_n,\bQ_{n'};-\bk-\bQ_{P-1}-\bq,\bq) =
(s^a)^n (s^b)^{n'} s^{ab} A_{\ell\ell'}^{ab}(\bQ_n,\bQ_{n'};\bk,\bq) \, ,
\end{equation}
with $s^{ab} = s^a s^b p^a p^b$.

Using Eq.~\eqref{eq: A sym} and the relation $E_{-\bk-\bQ_{P-1}}^\ell = E_\bk^\ell$, we can derive the following relation for the susceptibility, extending a derivation for spiral states \cite{Bonetti2022} to stripe states.
\begin{eqnarray} \label{eq: chi0 sym}
 \chi_0^{ab}(\bQ_n,\bQ_{n'};\bq,\omega) &=&
 - \frac{1}{2P} \int_\bk \sum_{\ell,\ell'}
 A_{\ell\ell'}^{ab}(\bQ_n,\bQ_{n'};\bk,\bq) \,
 \frac{f(E_\bk^\ell) - f(E_{\bk+\bq}^{\ell'})}
 {E_\bk^\ell - E_{\bk+\bq}^{\ell'} + \omega + i0^+} \nonumber \\
 &=& - \frac{1}{2P} \int_\bk \sum_{\ell,\ell'}
 \Big[ A_{\ell\ell'}^{ab}(\bQ_n,\bQ_{n'};\bk,\bq) \,
 \frac{f(E_\bk^\ell)}{E_\bk^\ell - E_{\bk+\bq}^{\ell'} + \omega + i0^+} \nonumber \\
 && - A_{\ell'\ell}^{ab}(\bQ_n,\bQ_{n'};-\bk\!-\!\bQ_{P-1}\!-\!\bq,\bq)
 \frac{f(E_{-\bk-\bQ_{P-1}}^\ell)}
 {E_{-\bk-\bQ_{P-1}-\bq}^{\ell'} - E_{-\bk-\bQ_{P-1}}^\ell + \omega + i0^+} \Big]
 \nonumber \\
 &=& - \frac{1}{2P} \int_\bk \sum_{\ell,\ell'}
 A_{\ell\ell'}^{ab}(\bQ_n,\bQ_{n'};\bk,\bq) \Big[
 \frac{f(E_\bk^\ell)}{E_\bk^\ell - E_{\bk+\bq}^{\ell'} + \omega + i0^+} \nonumber \\
 && + (s^a)^n (s^b)^{n'} s^{ab}
 \frac{f(E_\bk^\ell)}{E_\bk^\ell - E_{\bk+\bq}^{\ell'} - \omega - i0^+} \Big] \, .
\end{eqnarray}
In the second step we have exchanged the variables $\ell$ and $\ell'$ in the second term.
Denoting the contribution from ${\rm Re} F_{\ell\ell'}$ to $\chi_0^{ab}$ as $\chi_{0,R}^{ab}$, and the contribution from ${\rm Im} F_{\ell\ell'}$ as $\chi_{0,I}^{ab}$, we then obtain the following symmetry relations for a frequency sign change
\begin{subequations} \label{eq: chi0 freq sym}
\begin{align}
 \chi_{0,R}^{ab}(\bQ_n,\bQ_{n'};\bq,\omega) &=
 (s^a)^n (s^b)^{n'} s^{ab} \chi_{0,R}^{ab}(\bQ_n,\bQ_{n'};\bq,-\omega) \, , \\
 \chi_{0,I}^{ab}(\bQ_n,\bQ_{n'};\bq,\omega) &=
 - (s^a)^n (s^b)^{n'} s^{ab} \chi_{0,I}^{ab}(\bQ_n,\bQ_{n'};\bq,-\omega) \, .
\end{align}
\end{subequations}
These relations imply that $\chi_{0,R}^{ab}(\bQ_n,\bQ_{n'};\bq,0)$ and $\chi_{0,I}^{ab}(\bQ_n,\bQ_{n'};\bq,0)$ vanish for certain choices of $a$, $b$, $n$, and $n'$.
For example, $\chi_{0,R}^{aa}(\bQ_n,\bQ_{n'};\bq,0)$ vanishes for $a \in \{2,3\}$ if $n-n'$ is odd, while $\chi_{0,I}^{aa}(\bQ_n,\bQ_{n'};\bq,0)$ always vanishes for $a \in \{0,1\}$, and for $a \in \{2,3\}$ if $n-n'$ is even.

To derive symmetry relations for a momentum inversion $\bq \mapsto -\bq$, we shift the momentum variable $\bk$ in Eq.~\eqref{eq: chi0 a} by $-\bq/2$. Using the cyclic property of the trace, and the identities $\Pi_n^T = \Pi_{P-n}$ and
$\Pi_n^T \Sigma^a \Pi_n =  \Pi_n \Sigma^a \Pi_n^T = (s^a)^n \Sigma^a$,
one obtains the relation
\begin{equation}
 A_{\ell\ell'}^{ab}(\bQ_n,\bQ_{n'};\bk-\bq/2,\bq) =
 (s^a)^n (s^b)^{n'} A_{\ell'\ell}^{ba}(-\bQ_{n'},-\bQ_n;\bk+\bq/2,-\bq) \, .
\end{equation}
Note that $\bQ_{P-n} = - \bQ_n$. Combining this with Eq.~\eqref{eq: A*3} yields
\begin{equation} \label{eq: A parity}
 A_{\ell\ell'}^{ab}(\bQ_n,\bQ_{n'};\bk-\bq/2,\bq) =
 (s^a)^n (s^b)^{n'} p^a p^b A_{\ell'\ell}^{ab}(-\bQ_n,-\bQ_{n'};\bk+\bq/2,-\bq) \, .
\end{equation}
Inserting this relation into Eq.~\eqref{eq: chi0 a} with $\bk$ shifted by $-\bq/2$ yields
\begin{subequations}
\begin{align}
 \chi_{0,R}^{ab}(\bQ_n,\bQ_{n'};\bq,\omega) &=
 (s^a)^n (s^b)^{n'} p^a p^b \chi_{0,R}^{ab}(-\bQ_n,-\bQ_{n'};-\bq,-\omega) \, , \\
 \chi_{0,I}^{ab}(\bQ_n,\bQ_{n'};\bq,\omega) &=
 - (s^a)^n (s^b)^{n'} p^a p^b \chi_{0,I}^{ab}(-\bQ_n,-\bQ_{n'};-\bq,-\omega) \, .
\end{align}
\end{subequations}
Combining this with Eq.~\eqref{eq: chi0 freq sym} we obtain the symmetry relation for momentum inversion
\begin{equation}
 \chi_0^{ab}(\bQ_n,\bQ_{n'};\bq,\omega) =
 s^a s^b \chi_0^{ab}(-\bQ_n,-\bQ_{n'};-\bq,\omega) \, .
\end{equation}
Hence, for $a \in \{0,1\}$ and $b \in \{2,3\}$ (or vice versa),
$\chi_0^{ab}(\bQ_n,\bQ_{n'};\mathbf{0},\omega)$ vanishes if $n$ and $n'$ are even.


\bibliography{stripe.bib}

\end{document}